\documentclass[twocolumn]{aastex631}
\usepackage{amsmath}

\begin{document}

\title{Primordial Black Hole Triggered Type Ia Supernovae I: Impact on Explosion Dynamics and Light Curves}

\shortauthors{Leung, Walther, Nomoto and Kusenko}
\shorttitle{PBH Triggered Type Ia Supernovae I}

\author[0000-0002-4972-3803]{Shing-Chi Leung}

\affiliation{Department of Mathematics and Physics, SUNY Polytechnic Institute, 100 Seymour Road, Utica, NY 13502, USA}

\author[0009-0006-8467-2163]{Seth Walther}

\affiliation{Department of Electrical and Computer Engineering, SUNY Polytechnic Institute, 100 Seymour Road, Utica, NY 13502, USA}

\affiliation{Department of Mathematics and Physics, SUNY Polytechnic Institute, 100 Seymour Road, Utica, NY 13502, USA}

\author[0000-0001-9553-0685]{Ken'ichi Nomoto}

\affiliation{Kavli Institute for the Physics and 
Mathematics of the Universe (WPI), The University 
of Tokyo Institutes for Advanced Study, The 
University of Tokyo, Kashiwa, Chiba 277-8583, Japan}

\author[0000-0002-8619-1260]{Alexander Kusenko}

\affiliation{Department of Physics and Astronomy, University of California, Los Angeles California, 90095-1547, USA}

\affiliation{Kavli Institute for the Physics and 
Mathematics of the Universe (WPI), The University 
of Tokyo Institutes for Advanced Study, The 
University of Tokyo, Kashiwa, Chiba 277-8583, Japan}

\correspondingauthor{Shing-Chi Leung}
\email{leungs@sunypoly.edu}

\date{\today}

\submitjournal{ApJ}
\received{Jun 12 2025}
\revised{Jul 16 2025}
\accepted{Jul 25 2025}
\published{Sep 11 2025}

\newcommand{\red}[1]{\textcolor{red}{#1}}

\begin{abstract}

Primordial black holes (PBHs) in the asteroid-mass window  are compelling dark matter candidates, made plausible by the existence of  black holes and by the variety of mechanisms of their production in the early universe.  If a PBH falls into a white dwarf (WD), the strong tidal forces can generate enough heat to trigger a  thermonuclear runaway explosion, depending on the WD mass and the PBH orbital parameters. In this work, we investigate the WD explosion triggered by the passage of PBH. We perform 2D simulations of the WD undergoing thermonuclear explosion in this scenario, with the predicted ignition site as the parameter assuming the deflagration-detonation transition model. We study the explosion dynamics and predict the associated light curves and nucleosynthesis. We find that the model sequence predicts the light curves which align with the Phillip's relation ($B_{\max}$ vs. $\Delta M_{15}$). Our models hint at a unifying approach in triggering Type Ia supernovae without involving two distinctive evolutionary tracks.

\end{abstract}





\section{Introduction}
Primordial black holes could have formed in the dense plasma of the early universe long before the stars and galaxies are born~\citep{Zeldovich:1967,Hawking:1971ei,Carr:1974nx}. In the asteroid-mass window
($4 \times 10^{-17} \lesssim M_{\rm PBH} \lesssim 4 \times 10^{-12}~M_{\odot}$), PBHs could account for all dark matter.
The preponderance of possibilities for their formation and the fact that black holes are known to exist in nature makes PBH a particularly plausible dark-matter candidate~\citep{Green:2020jor,Carr:2020xqk}.  There are also observational hints for small PBHs in microlensing, r-process data, and other astrophysical data~\citep{Niikura:2017zjd,Fuller:2017uyd,Carr:2019kxo,Takhistov:2020vxs,Coogan:2020tuf,Sugiyama:2021xqg,Lu:2023xoi,Carr:2023tpt,Flores:2023lll}.  If a PBH falls into a white dwarf (WD), it can trigger a thermonuclear runaway explosion, leading to a Type Ia supernova \citep{Graham2015DMSNIa}.

\subsection{Type Ia Supernova Diversity}

Type Ia supernovae (SNe Ia) are the explosions of carbon-oxygen white dwarfs (C+O WDs). The observed spectroscopic and photometric diversity has inspired a wide range of explosion models. How SNe Ia are triggered and formed remains a matter of debate \citep[see recent models in ][]{Hillebrandt2000SNIaReview, Nomoto2018SNIa, Leung2023SNIaReview}. Major candidates include (1) the Chandrasekhar mass WD \citep{Nomoto1984SNIaIII}, the sub-Chandrasekhar mass WD \citep[e.g.,][]{Fink2007SubChand}, and their variations including the pulsation reverse detonation model \citep{Bravo2009PulsatingI, Bravo2009PulsatingII}, the direct collision model \citep{Rosswog2009WDCollision} and the gravitationally confined detonation model \citep{Plewa2004GCD}.  If a PBH falls into a white dwarf (WD), it can trigger a thermonuclear runaway explosion, leading to a Type Ia supernova \citep{Graham2015DMSNIa}.

In the Chandrasekhar mass model (Ch-mass), it relies on the WD accreting matter from its evolved main-sequence companion star (primarily single degenerate scenario, SD). Mass transfer continues until the matter becomes degenerate. At that point, the nuclear burning becomes unstable and develops the first runaway through the sub-sonic deflagration \citep{Niemeyer1995TurbulentFlame}. When the turbulent deflagration enters the distributed regime, it transits to super-sonic detonation \citep{Khokhlov1991DDT,Brooker2021DDTCrit} which consumes the entire star \citep{Roepke2007DDT, Leung2018Chand}. The univque Ch-mass of the WD supports the observed similarities in SNe Ia. Meanwhile, its variation in the initial runaway geometry can seed for the diversity \citep{Seitenzahl2013DDT}. Some SNe Ia feature the H-rich circumstellar medium which agrees with the mass accretion history of the SD scenario. However, shortcoming exists in the SD scenario including that (1) the explosion should result in a UV-flash \citep{Liu2015UVFlash}, but is rarely observed except for the case iPTF14atg \citep{Kromer2016iPTF14atg}\footnote{With another caveat that the spectral evolution of the SD model does not fit the data}; and (2) the thermonuclear runaway can be triggered only for a limited parameter space for the mass accretion: A too high or low accretion rate results in a nova, double detonation or direct collapse \citep{Nomoto1982SNIaII}. 

The sub-Chandrasekhar mass (subCh-mass) channel relies on external triggers, e.g., the detonating He shell \citep[e.g., ][]{Nomoto1982SNIaII, Sim2010SubChand, Moll2013SubChand, Leung2020SubChand} or the violent merger process \citep[e.g., ][]{Pakmor2012Merger} in binary WD systems. The explosion can be evolved in both SD and double degenerate (DD) scenarios. In both pictures, the unstable He-burning creates the hot spot in the He-envelope \citep{Tanikawa2019Merger}, which develops into detonation, and propagates into the core. The collision and the WD disruption can naturally explain the unobserved companion stars in the SN Ia environment \citep[see, e.g., ][]{Shields2023NoSNIaCompanion}. The collision process avoids the missing long-term soft X-ray signals in the SD scenario. However, from the chemical perspective, the sub-Chandrasekhar mass WD does not have the required high density to support electron-capture and the low-$Y_e$ isotopes, which are indispensable for creating $^{55}$Co, the seed for $^{55}$Mn \citep{Seitenzahl2013Mn, Nomoto2017HBSN}. Also, the asymmetry in the He-/C-envelope due to pre-collision turbulent motion may fail to induce second (C-)detonation \citep{Fenn2016}.

\subsection{Phillips Relation and Motivation}

SNe Ia are unique because of the Phillips relation \citep{Phillips1999Cosmo}. Their peak brightness of the B-band correlates with their decline rate. Ordinary SN Ia light curves (LCs) can be categorized by a single parameter family, which provides a powerful tool for deriving their absolute brightness by their measured width \citep{Perlmutter1999Cosmo}. 

A challenge in SN Ia modeling is how to reconcile explosion models with the observed similarity. Multiple works have studied how the model parameter could create SN Ia model sequences that match the Phillips's relation, including the central density \cite[e.g., ][]{Ohlmann2014} for the Ch-mass models, and the progenitor mass in the subCh-mass models \citep{Woosley2007SubChand, Sim2013Phillips, Shen2021SubChand}.

Therefore, both Ch-mass and subCh-mass models for SNe Ia have their individual importance in the chemical and optical perspective. The two channels have their evidence and limitations to their formation picture. It becomes interesting to study if there is a unifying picture that can facilitate the explosion of both progenitors in the nature. In fact, the accretion of primordial black hole \citep[PBH, ][]{Hawking1971PBH, Carr1975} of the asteroid size in a white dwarf is shown feasible triggering the initial hot spot for both progenitor types \citep[MC19, ][]{Montero2019PBH}. 

PBH is one of the possible candidates for cosmological dark matter. Searches for microlensing results from various surveys, including MACHO \citep[Massive Compact Halo Objects, ][]{Alcock2001MACHO}, EROS \citep[Expérience pour la Recherche d'Objets Sombres, ][]{Tisserand2007EROS}, OGLE-5 \citep[Optical Gravitational Lensing Experiment, ][]{Niikura2019OGLE5} and that by the Kepler Space Telescope \citep{Griest2014Kepler}, have narrowed the mass fraction of PBH with a mass between $10^{-11} M_{\odot}$ to a few $M_{\odot}$ as a DM candidate down to a few percent. PBH in the astroid mass (below $10^{-11}~M_{\odot}$) is not constrained and could fully explain the entire dark matter mass budget. The Schwarzschild radius for these asteroid-mass PBHs can be as small as a few nanometer. 

MC19 studied the dynamical response of the WD when such PBHs enter by one-dimensional microscopic simulations. The strong but local gravitational force of the PBH attracts C+O-rich matter in the vicinity, where the compressional heating forms a hot spot along the PBH trajectory in the WD. Depending on the initial angular momentum of the PBH, the orbital motion creates a bubble-like or a ring-like hot zone. They presented the range of WD mass and impact parameter where the hot spots can develop into thermonuclear runaway inside the WD. The WD progenitor mass range widely from subCh-mass to Ch-mass. Thus this framework provides a unifying framework that bridges the two distinctive classes in the same footing, with the WD mass being the only parameter to be varied. It becomes interesting to study, when the SN Ia is triggered by this mechanism, what diversity of models it can associate with, and where they correspond to in the Phillip's relation. 

In this article, we study the explosion, its nucleosynthesis and radiative transfer of the SNe Ia in this physical picture. In Section \ref{sec:method} we present our numerical tools used in this work. In Section \ref{sec:characteristic}, we present our characteristic model where the explosion resembles with typical SN Ia explosion ($^{56}$Ni mass about 0.5 -- 0.7 $M_{\odot}$). Then in Section \ref{sec:dependence} we explore how the explosion varies along the sequence of models predicted in MC19, focusing on the light curves and its implied parameter space in the Phillips relation. At last in Section \ref{sec:discussion}, we compare our results with other SN Ia explosion models in the literature and discuss the possible extension of this project. At last, we give our conclusion. 

\section{Method}
\label{sec:method}

\subsection{Numerical Simulations}

\begin{table*}[]
    \centering
    \caption{The initial models and the global parameters of models presented in this article. $M$, $M_{\rm ign}$, and $M_{\rm Ni}$ are the initial model mass, mass coordinate of the initial ignition and the total synthesized $^{56}$Ni at the end of simulations, in units of $M_{\odot}$. The variable $m_{\rm ign}$ is the initial ignited mass in deflagration in units of $10^{-4} M_{\odot}$. $R$ and $R_{\rm ign}$ are the initial model radius and the radial position of the initial ignition, in units of km. $E_{\rm exp}$ is the final energy of the ejecta, in units of $10^{50}$ erg. The variables $\rho_c$ and $\rho_{\rm ign}$ are the central and ignition density of the initial WD model in g cm$^{-3}$. [X/Fe] is the elemental mass fraction scaled with the solar value, defined as [X/Fe] $= \log_{10} $(X/X(Fe))/(X/X(Fe))$_{\odot}$. The column "Shape" represents the initial thermonuclear runaway geometry. The column "others" acts as a remark on whether the models are selected with Kelvin-Helmholtz instabilities, derived from Fig. 7 of MC19. The model name follows the format XXY-Z, where XX is the mass coordinate of the initial ignition in units of 0.1 $M_{\odot}$; Y is the initial ignition geometry with `R' being a ring and `B` being a bubble; an optional tag Z marks the models where Kelvin-Helmholtz (KH) instabilities are insignificant in the heating phase.}
    \begin{tabular} {c c c c c c c c c c c c c c}
        \hline
         Model & $M$ & $\rho_c$ & $R$ & $M_{\rm ign}$ & $\rho_{\rm ign}$ & $R_{\rm ign}$ & $m_{\rm ign}$ & $E_{\rm exp}$ & $M_{\rm Ni}$ & [Mn/Fe] & [Ni/Fe] & Shape & Others \\
         \hline
         02B-noKH & 0.95 & $2.35\times 10^{7}$ & 6240 & 0.2 & $1.31 \times 10^{7}$ & 1780 & 0.11 & 9.27 & 0.297 & 0.562 & 0.825 & bubble & no KH \\
         04B-noKH & 1.02 & $3.94\times 10^{7}$ & 5600 & 0.4 & $1.33 \times 10^{7}$ & 2103 & 0.10 & 11.48 & 0.504 & 0.314 & 0.912 & bubble & no KH \\
         06B-noKH & 1.08 & $5.62\times 10^{7}$ & 5200 & 0.6 & $1.28 \times 10^{7}$ & 2280 & 0.10 & 12.35 & 0.586 & 0.247 & 0.910 & bubble & no KH \\
         08B-noKH & 1.15 & $9.83\times 10^{7}$ & 4570 & 0.8 & $1.31 \times 10^{7}$ & 2290 & 0.11 & 13.70 & 0.770 & 0.160 & 0.948 & bubble & no KH \\
         10B-noKH & 1.26 & $3.11\times 10^{8}$ & 3430 & 1.0 & $2.30 \times 10^{7}$ & 1860 & 0.02 & 14.54 & 1.032 & 0.295 & 1.050 & bubble & no KH \\
         02B & 1.21 & $1.74\times 10^{8}$ & 3930 & 0.2 & $1.02 \times 10^{8}$ & 910 & 0.10 & 13.74 & 0.952 & 0.081 & 1.078 & bubble & with KH \\
         06B & 1.28 & $4.10\times 10^{8}$ & 3210 & 0.6 & $1.05 \times 10^{8}$ & 1150 & 0.10 & 16.38 & 1.057 & 0.475 & 1.014 & bubble & with KH \\
         10B & 1.34 & $1.25\times 10^{9}$ & 2430 & 1.0 & $1.06 \times 10^{8}$ & 1150 & 0.11 & 16.87 & 1.084 & 1.208 & 1.245 & bubble & with KH \\
         \hline
         02R-noKH & 0.95 & $2.35\times 10^{7}$ & 6240 & 0.2 & $1.31 \times 10^{7}$ & 1780 & 5.47 & 10.62 & 0.403 & 0.418 & 1.030 & ring & no KH \\
         04R-noKH & 1.02 & $3.94\times 10^{7}$ & 5600 & 0.4 & $1.33 \times 10^{7}$ & 2103 & 7.17 & 10.62 & 0.576 & 0.222 & 1.038 & ring & no KH \\
         06R-noKH & 1.08 & $5.62\times 10^{7}$ & 5200 & 0.6 & $1.28 \times 10^{7}$ & 2280 & 7.22 & 12.60 & 0.655 & 0.171 & 1.042 & ring & no KH \\
         08R-noKH & 1.15 & $9.83\times 10^{7}$ & 4570 & 0.8 & $1.31 \times 10^{7}$ & 2290 & 7.98 & 13.70 & 0.816 & 0.120 & 1.099 & ring & no KH \\
         10R-noKH & 1.26 & $3.11\times 10^{8}$ & 3430 & 1.0 & $2.30 \times 10^{7}$ & 1860 & 2.67 & 16.07 & 1.04 & 0.289 & 1.219 & ring & no KH \\
         02R & 1.21 & $1.74\times 10^{8}$ & 3930 & 0.2 & $1.02 \times 10^{8}$ & 910 & 5.75 & 16.74 & 1.01 & 0.081 & 1.078 & ring & with KH \\
         06R & 1.28 & $4.10\times 10^{8}$ & 3210 & 0.6 & $1.05 \times 10^{8}$ & 1150 & 7.55 & 17.64 & 1.08 & 0.373 & 1.258 & ring & with KH \\
         10R & 1.34 & $1.25\times 10^{9}$ & 2430 & 1.0 & $1.06 \times 10^{8}$ & 1150 & 8.01 & 17.28 & 1.07 & 0.944 & 1.419 & ring & with KH \\
         \hline
    \end{tabular}
    \label{tab:init_model}
\end{table*}

We use the 2D hydrodynamics code which solves the Euler equations in cylindrical coordinates reported in \cite{Leung2015WENO}. The code uses the fifth order Weighted Essentially Non-oscillatory shock capturing scheme for the space-discretization \citep{Barth1999} and the five-step third-order non-strong stability-preserving Runge-Kutta scheme for the time discretization \citep{Wang2007WENO}. 

We model the explosion by the deflagration-detonation transition with turbulent deflagration. The sub-grid turbulence follows the prescription in \cite{Niemeyer1995TurbulentFlame}. The code uses the '\texttt{Helmholtz} equation of state (EOS) \citep{Timmes1999Helmholtz, Timmes2000Helmholtz}, which provides the thermodynamics state from a given density $\rho$, temperature $T$ and atomic mass $A$ and atomic number $Z$. In the hydrodynamics simulations, the chemical composition is represented by the 7-isotope network \citep{Timmes20007Iso}. To capture the nuclear runaway front, we use the level-set method prescribed in \cite{Reinecke1999LevelSet} with the laminar and turbulent flame physics described in \cite{Timmes1992COFLame, Hicks2015RTFlame}.

The code has been used extensively in our modeling of SNe Ia including the Chandrasekhar mass WD assuming DDT \citep{Leung2018Chand}, PTD \citep{Leung2020SNIax} and sub-Chandrasekhar mass WD \citep{Leung2020SubChand}. The SN Ia models have been applied to explain the observed light curves of recent supernovae \citep{Tiwari2022SNIaLC, Leung2021SN2014J}, the observed chemical abundance of supernova remnants \citep{Zhou2021SgrAE,Weng2022SNR} and the chemical abundances of halo stars in the Milky Way Galaxy \citep{Ishigaki2021MPS}.

In the simulations, the level-set method captures the geometry of both deflagration and detonation wave fronts. When the front sweeps the C+O-rich matter, energy is injected based on the mass of burnt region. Combusted matter first reaches the nuclear quasi-statistical equilibrium burning, which forms Si-group elements peaked at $^{28}$Si. Then, for matter with a density $\rho \geqslant 5 \times 10^7$ g cm$^{-3}$, the burning proceeds until it ends in $^{56}$Ni. We use the nuclear burning timescale \citep[][]{Calder2007} to model the slow O- and Si-burning into NSE. For completely incinerated zones, when the local temperature exceeds $\sim 5\times 10^9$ K, nuclear statistical equilibrium (NSE) and electron capture take place where the chemical composition follows the pre-computed composition as a a function of $(\rho, T, Y_e)$ \citep{Reinecke1999SNIa, Nomoto2017HBSN}. The temperature is corrected such that the compositional and energy change  due to NSE is consistently included. In general, the NSE acts as an energy reservoir: when temperature change due to hydrodynamics is buffered by the change of binding energy for the combination and photo-disintegration between $^{56}$Ni and $^4$He. 

We use the passive tracer particles scheme to do the post-process nucleosynthesis> It captures the thermodynamics history of individual Lagrangian fluid elements \citep{Travaglio2004Tracer}. The tracers only follow the Eulerian fluid motion, and record the local density and temperature as a function of time. The thermodynamics history is then post-processed by a 495-isotope nuclear reaction network \texttt{Torch} \citep{Timmes1999Torch} to compute the detailed explosive nucleosynthesis. In each simulation, $160^2$ Lagrangian tracers are placed to represent the star in equal-mass portions. We refer interested readers to \cite{Seitenzahl2010Tracer} for the convergence study of post-processing nucleosynthesis using tracer particles, with uncertainty about a few \% for our tracer count.

After the hydrodynamics and nucleosynthesis, we pass the models to the radiative transfer code \texttt{SNEC} \citep[Supernova Explosion Code, ][]{Morozova2015SNEC} to solve for the bolometric light curves. The code is developed based on the prototype reported in \cite{Bersten2011SNEC, Bersten2013SNEC}. The code solves the radiative hydrodynamics assuming the blackbody diffusion approximation. The code models the post-explosion phase and extracts optical signals from the ejecta until it becomes transparent. It computes the light curve, kinematics of the photosphere and the individual photometry band magnitude. 

We refer interested readers to our documentation article for the code tests and applications. 

\subsection{Initial Models}

The initial models are constructed as isothermal C+O white dwarfs in hydrostatic equilibrium, with a given progenitor mass $M$. Then, we trigger the initial nuclear runaway by putting an initial ignition site at a given radius $R_{\rm ign}$ and mass coordinate $M_{\rm ign}$, obtained by interpolation from Figure 7 of MC19. In that article, a range of $M_{\rm ign}$ is provided for a given $M$ where such ignition is possible. We list in Table \ref{tab:init_model} the initial setting of the C+O white dwarf models and the location where the initial nuclear runaway takes place. Each model is denoted in the form XXY-Z. XX is the the ignition mass coordinate in 0.1 $M_{\odot}$. Y stands for the initial runaway geometry, which can be a ring `R' or a bubble `B'. Z stands for whether the model corresponds to the series where the KH instability is important (default with KH instability). Models without the Z-label refer to those where KH instability is significant during the PBH heating phase (noted as the KH series). Models where KH instability is less important to the PBH heating is noted as the noKH series. 

In MC19, they have classified the exploding models into two scenarios, whether KH instability is significant to the cooling of the PBH induced heating. During the orbital motion of the PBH inside the WD, the KH instability can facilitate mixing of cold matter and remove the thermal energy generated by nuclear fusion. Models where KH instability is ineffective can accumulate the heat and ignite at a lower density, thus resulting in a lower progenitor mass for a given ignition mass.  All initial nuclear deflagration is triggered by hand at the equator plane (x-axis) or along the rotation axis (z-axis). The orbital plane of the PBH defines the equatorial plane. 

In general, the PBH passes through the WD in a straight line where the nuclear runaway occurs along the trail. The actual structure of the nuclear runaway depends on the dynamical friction of the PBH orbiting inside the WD. In this work, we study two extremes. If the dynamical friction is small, the gravity of the PBHs generates a hot bubble along its orbit, which, in the minimal case, corresponds to the highest density location. The resulting ignition is similar to a ``bubble'' shape as a single-spot ignition. On another extreme, if the dynamical friction is strong and the initial energy and angular momentum of the PBH are low, the PBH orbit could converge to a closed orbit. During its orbital motion, the sustained burning along the orbit triggers the initial runaway in the form of a ring. Regardless of noKH and KH, in the cylindrical coordinate with rotation symmetry, they are effectively a semi-circle along either of the axis. To avoid artificially exciting the WD, where the matter is highly degenerate, we place a small initial runaway to launch the explosion. The initial runaway appears as a semicircle bubble of $\sim 2$ grids in radius. This corresponds to $\sim 10^{-5}~M_{\odot}$ ($10^{-4}~M_{\odot}$) along the symmetry (equator) plane. The similarity of the results for the ring or bubble implies that the initial explosion seed does not significantly affects the later explosion dynamics. In Appendix B we add a comparative study to show that our results are not sensitive to $m_{\rm ign}$.

A consistent trend can be observed in the initial models. With the same $M_{\rm ign}$, tho no-KH series requires a lower progenitor mass than the models in the KH series: The shear instabilities create turbulent flow and mixing, which dissipates heat from the PBH. The resulting WD has a larger radius $R$ and a lower initial central density $\rho_c$. The ignition density $\rho_{\rm ign}$ is also substantially lower where the runaway commences at a more exterior mass coordinate. 

By comparing models of the KH-series, a higher $M_{\rm ign}$ corresponds to a higher mass WD, much closer to the Chandrasekhar mass. The models have higher $\rho_c$ and lower $R$. There is no significant change in the ignition density or their actual location. 

\subsection{Pre-Explosion WD Models}
\label{sec:preexp}

\begin{figure}
    \centering
    \includegraphics[width=8cm]{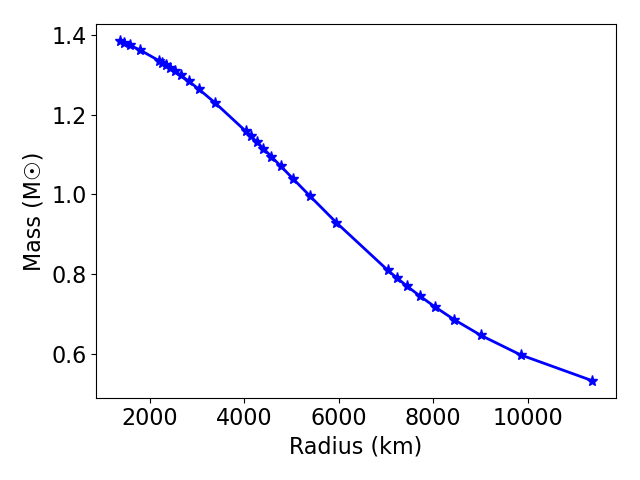}
    \caption{The mass-radius of the WD models with each data point marked by an asterisk. }
    \label{fig:32models}
\end{figure}

\begin{figure}
    \centering
    \includegraphics[width=10cm]{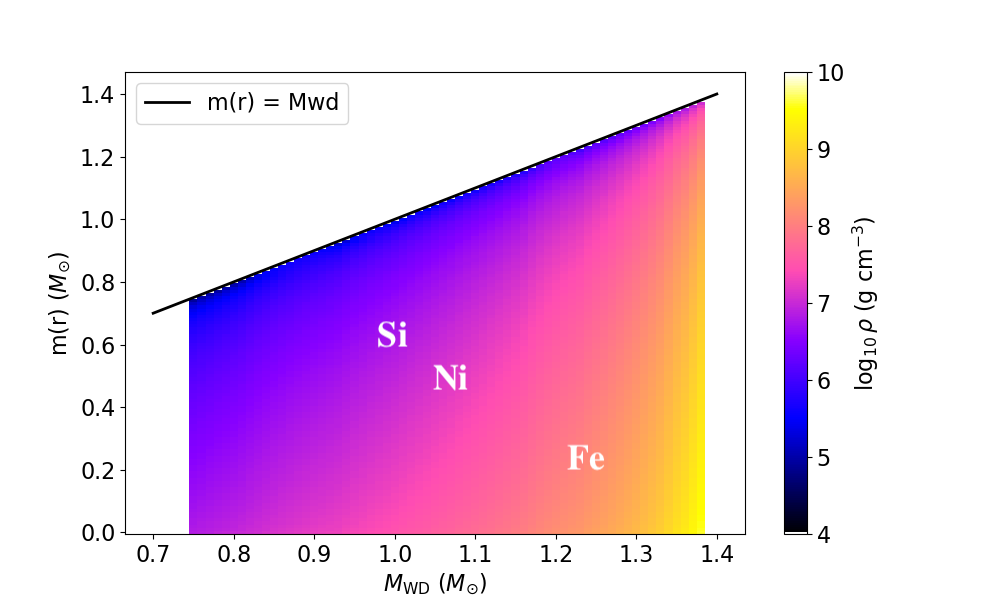}
    \caption{The density colour plot for WD models of various masses. The expected final composition after self-heating nuclear reaction are indicated by the dominant elements: Si, Ni, and Fe.}
    \label{fig:dens_profile_model01}
\end{figure}


To locate the mass and radial coordinates of the ignition positions, we use the fourth-order Runge-Kutta method to solve for the hydrostatic equations of the isothermal WD models.  We construct 32 carbon-oxygen WD models using the \texttt{Helmholtz} EOS \citep{Timmes2000Helmholtz}, for models with central densities from $1 \times 10^{6}$ to  $5 \times 10^{9}$ gcm$^{-3}$. In Figure \ref{fig:32models} we show the mass-radius relation of the hydrostatic models assuming $T = 10^8$ K, and $^{12}$C:$^{16}$O = 1:1 in mass fraction. 

To estimate the nucleosynthetic yields, we calculate in each WD model the thermonuclear reaction at regular mass coordinate intervals by using the 21-isotopic network \citep{Timmes1999Torch}. This lets us estimate how robust of the nuclear burning in each location. We assume the nuclear burning takes place at a constant density. The 21-isotope network provides an accurate estimate for the nuclear reactions of C+O rich matter along the $\alpha$-chain up to $^{56}$Ni. In Figure \ref{fig:dens_profile_model01}, we show how the density varies with the mass coordinate $m(r)$ in the initial WD for various WD masses $M_{\rm WD}$. We mark qualitatively the regions of the principal chemical elements produced based on the nuclear reaction network above. 

\begin{figure}
    \centering
    \includegraphics[width=10cm]{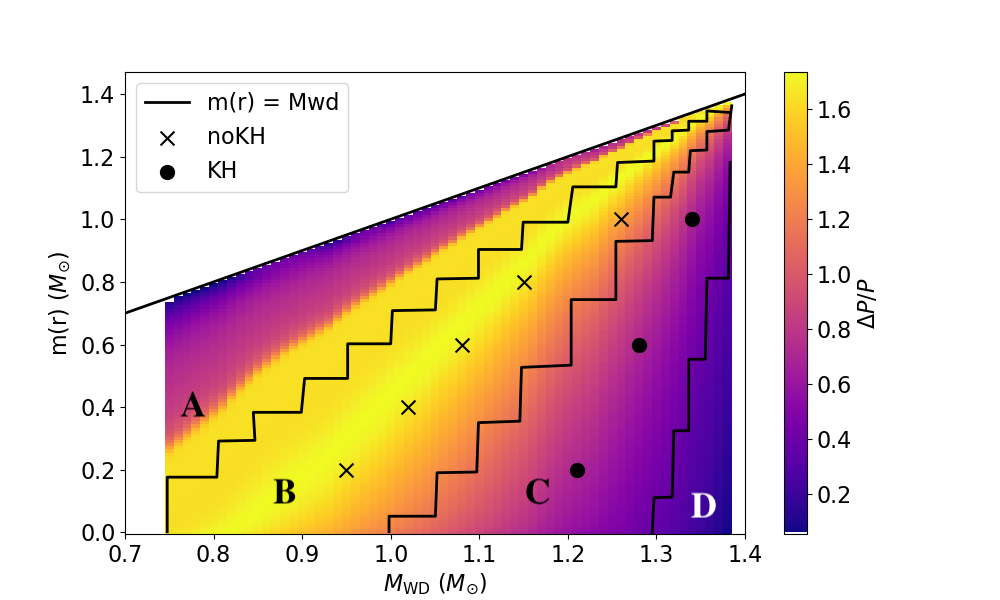}
    \includegraphics[width=10cm]{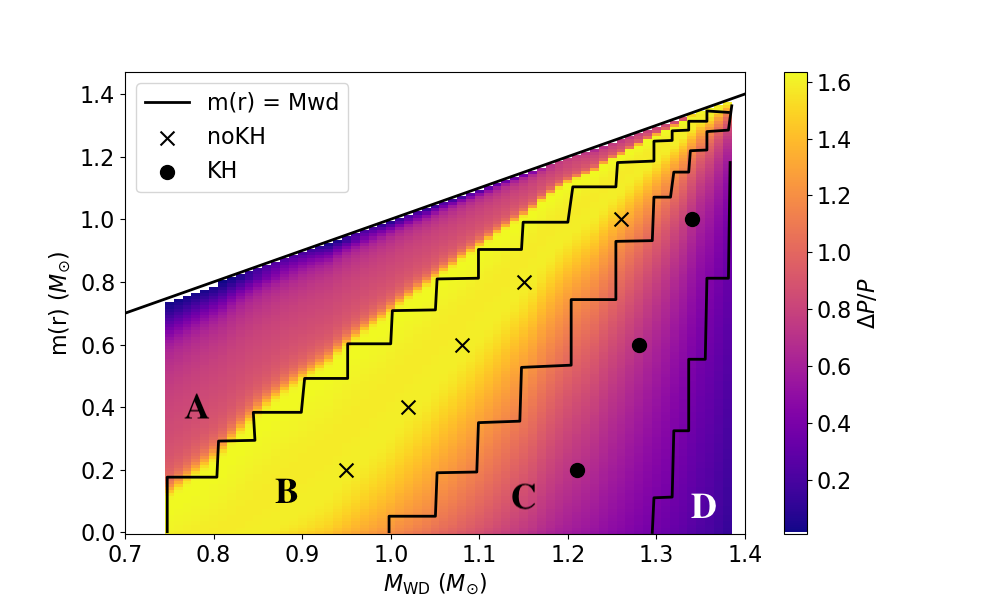}
    \caption{(top panel) Colour plot for the pressure jump after the nuclear burning taking place for 1 second. Region A, shows no nuclear runaway possible. Regions B (C) show the active regions for the cases where Kelvin-Helmholtz instability is insignificant (significant). Region D shows where carbon dissociation takes place due to extreme temperature. 
    (bottom panel) Same as top panel but with nuclear burning being 0.1 second.}
    \label{fig:1_second_pressure}
\end{figure}

To quantify whether the nuclear runaway appears in the form of deflagration or detonation, we follow \cite{Nomoto1982SNIaII} and compute the the pressure difference before and after the nuclear burning. In Figure \ref{fig:1_second_pressure} we show the results, assuming the nuclear burning takes place for 1 and 0.1 s respectively. We mark the regions following the scheme in MC19: Region A stands for no nuclear runaway being possible; Regions B (C) stand for nuclear runaway being possible when KH Instability is insignificant (significant). Region D stands for models where the carbon dissociation takes place due to a high temperature, and it is unclear if runaway is possible because nuclear statistical equilibrium and electron capture become important. The SN Ia models used in Table \ref{tab:init_model} are marked as crosses and circles in the figure. 

Both figures show clearly that the pressure jump with or without KH instability is comparable or much less than the pre-burning pressure $P_i$ of the matter. The highest value of $\sim 1.6 P_i$ occurs for matter supposed to burn into $^{56}$Ni. The small pressure jump suggests that the pressure gradient is unlikely to drive the detonation wave, and thus deflagration is the preferred channel. However, we remark that small-scale simulations are necessary to resolve the dynamics of high pressure zones.

In Appendix A, we show the energy production and representative atomic number of these models. 

\section{Characteristic Model}
\label{sec:characteristic}

\subsection{Characteristic Model of the noKH Series}

We first use one of the models, 06R-noKH, to illustrate how the explosion takes place and its typical energetics. The model is slightly subCh-mass (1.02 $M_{\odot}$) and the ignition takes place at an intermediate density ($\sim10^8$ g cm$^{-3}$). The explosion results in about 0.6 $M_{\odot}$ $^{56}$Ni in the ejecta, which agrees with typical SNe Ia for 0.5 -- 0.7 $M_{\odot}$ of $^{56}$Ni observed in the ejecta. In that perspective, 06R-noKH can be also viewed as the characteristic model.  

\begin{figure*}
    \centering
    \includegraphics[width=8.5cm]{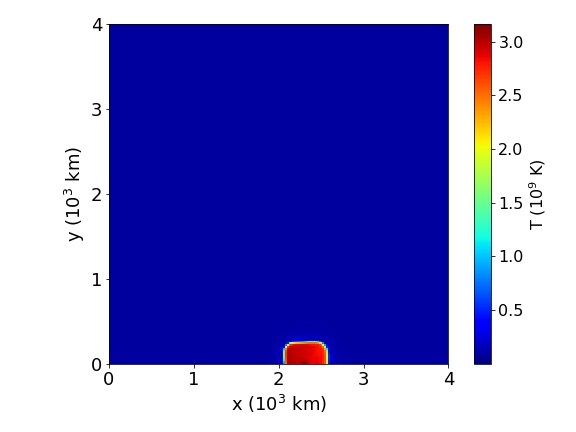}
    \includegraphics[width=8.5cm]{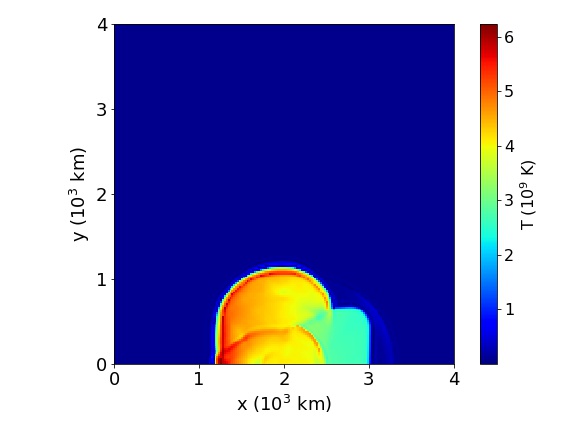}
    \includegraphics[width=8.5cm]{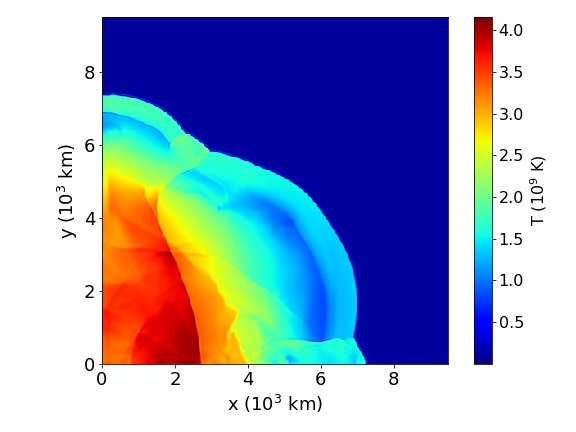}
    \includegraphics[width=8.5cm]{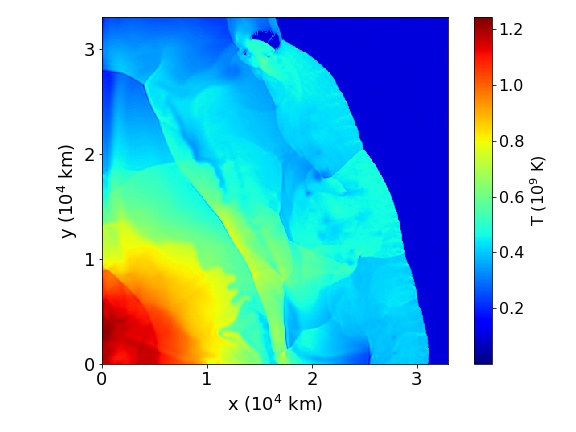}
    \caption{The temperature colour-map of the Model 06R-noKH taken at 0.1 (top left), 0.3 (top right), 0.57 (bottom left), and 1.15 s (bottom right), measured from the onset of ignition. Notice the length scale in the bottom right plot is in units of $10^4$ km.}
    \label{fig:temp_profile_model02}
\end{figure*}

In Figure \ref{fig:temp_profile_model02} we plot the temperature profile at 0.10, 0.30, 0.57 and 1.15 s, counted from the onset of ignition. At the beginning, the ``ring'' (which by symmetry is equivalent to a ``bubble'' on the equator) in the WD grows slowly as the ignition density is low compared to canonical Ch-mass SN Ia models. At 0.3 s, the ash has expanded to the outer half. The upper circular arc is the detonation front. Its larger size compared to the region burnt by deflagration reflects the higher propagation speed than the deflagration waves. At 0.57 s, the detonation has reached the surface where segment of hot zones $\sim$2000 km from center is the result of shock front collision. The outer surface has begun its expansion, while the ejecta remains spherical to a good approximation. Finally, at 1.15 s the entire WD is already incinerated. The central hot region appears less symmetric where relics of shock wave collision are visible. The `flurry' shape bear the $10^4$ km shell is a sign of Rayleigh-Taylor instability as the hot core expands into the cold outer layer. The maximum temperature sharply drops from $6 \times 10^9$ K to below $2 \times 10^9$ K, where most exothermic nuclear reactions have ceased. The detonation waves burn most of the WD. 

\begin{figure*}
    \centering
    \includegraphics[width=8.5cm]{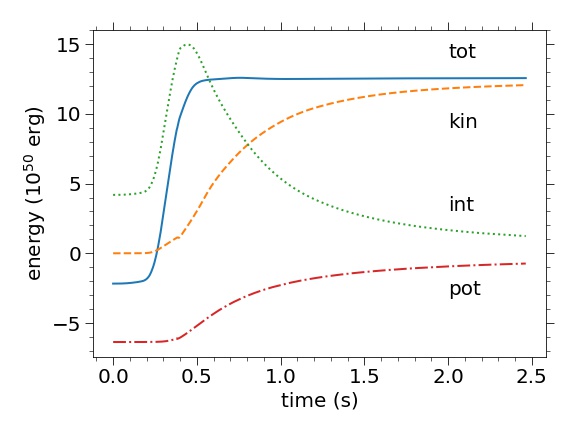}
    \includegraphics[width=8.5cm]{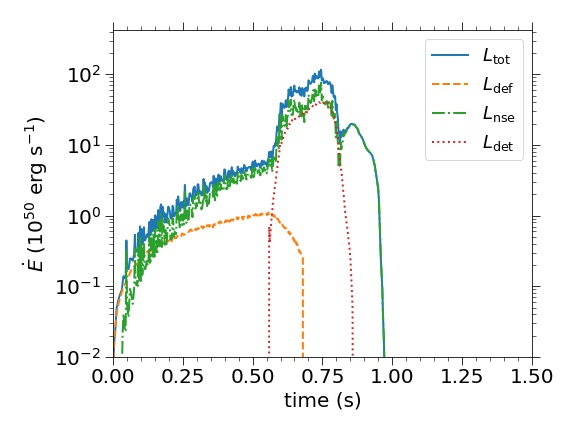}
    \caption{The energy evolution of the characteristic model including the total energy (tot; blue solid line), kinetic energy (kin; orange dashed line), internal energy (int; green dotted line) and the potential energy (pot; red dotted line).
    (right panel) The total energy production rate (blue solid line), and its components including the C-deflagration (orange dashed line), the NSE (green dot-dashed line) and the C-detonation (red dotted line) of the characteristic model.}
    \label{fig:model02wKH_energy}
\end{figure*}

In the left panel of Figure \ref{fig:model02wKH_energy} we plot the energy evolution of the characteristic model 06R-noKH. Most energy is produced between 0.5 -- 1.0 s. After that the model stays at a constant energy. During deflagration, energy is injected as internal energy, where the matter remains to a good approximation hydrostatic. The kinetic energy becomes more important in the energy budget in the next 1 second after the detonation transition. The internal energy sharply increases after the rise of the total energy, in the form of thermal energy. Following the expansion in the next 1 second, the pressure work done gradually converts internal energy into kinetic energy, which sustains the matter to escape from its self-gravity. Lastly, the internal energy contributes to $\sim10 \%$ of the total energy budget. Most energy of the ejecta is stored as kinetic energy. 

In the right panel, we plot the energy production history of the characteristic model. The model begins with C-deflagration, and the advanced burning follows immediately. The advanced burning brings in most of the energy. The deflagration itself contributes to about $10 \%$ of the energy generation. Unlike canonical SN Ia models, the transition to DDT happens at a much earlier time in this model ($\sim 0.6$ s). Once it takes place, the detonation releases more energy in a short time than deflagration. It consumes the remaining fuel within 0.3 s, bringing a sharp drop in the energy production. The expansion of the star triggers the NSE to release binding energy by recombining of $^{4}$He into $^{56}$Ni \cite[see e.g., ][]{Reinecke1999SNIa, Reinecke2002SNIa}, until all matter leaves NSE and matter becomes too cold for further recombination. 

\begin{figure}
    \centering
    \includegraphics[width=8.5cm]{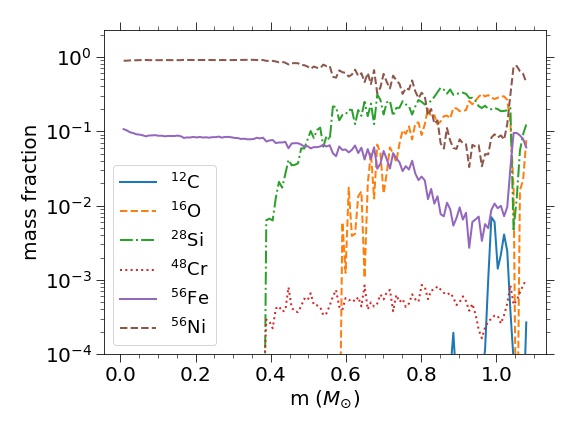}
    \includegraphics[width=8.5cm]{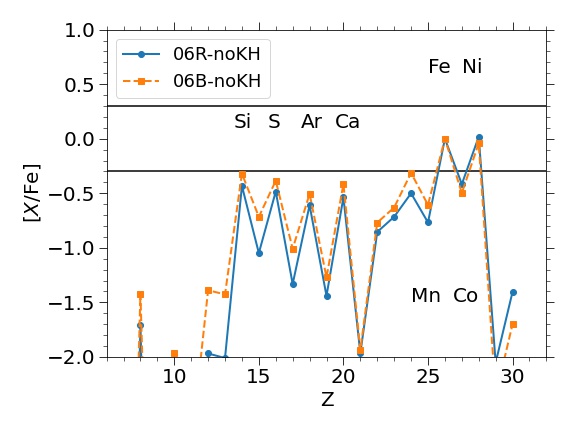}
    \caption{(top panel) The mass distribution of the ejecta for representative isotopes including $^{12}$C (blue solid line), $^{16}$O (orange dashed line), $^{28}$Si (green dot-dash line), $^{48}$Cr (red dotted line), $^{56}$Fe (purple solid line) and $^{56}$Ni (brown dashed line) in 06R-noKH.
    (bottom panel) The scaled mass fraction [X/Fe] $= \log_{10}$ [(X/Fe)/(X/Fe)$_{\odot}$]. of the characteristic models. The upper and lower horizontal lines stand for two times and half of the solar values. }
    \label{fig:model02wKH_abund}
\end{figure}

After the hydrodynamics simulations, the ejecta develops homologous expansion. We collect statistics of the tracer particles by their radial position, and construct their one-dimensional representation of the density, temperature and composition profile. In the top panel of Figure \ref{fig:model02wKH_abund} we plot the abundance of the selected elements from our post-processed tracer particles. The detonation makes most C+O matter be  completely incinerated and form $^{56}$Ni, indicated by the uniform $^{56}$Ni distribution in half of the star by mass. Near the surface, the drop of density leads to incomplete burning, and Si-group elements are produced. In the outermost 0.1 $M_{\odot}$, s of the fuel $^{16}$O persists in the ejecta.

In the bottom panel of Figure \ref{fig:model02wKH_abund} we plot the scaled mass fraction [X/Fe]\footnote{[X/Fe] = $\log_{10}$ (X/X(Fe))/(X/X(Fe))$_{\odot}$.} of the ejecta for 06R-noKH and 06B-noKH. The model exhibits a strong odd-even element parity in the ejecta. The complete burning leads to formation of Fe-group elements where the mass ratios of Fe and Ni are close to the solar composition. Most other elements that are representative in canonical SNe Ia, e.g., Si, S, are underproduced. Where the ignition is started (ring or bubble) does not significantly change the nucleosynthetic pattern. The two models show a similar chemical yield in different ignition kernels. Also, the subcCh-mass nature of the WD results in the sub-solar production of [Mn/Fe].

\subsection{Characteristic Model of the KH Series}

The importance of KH instability is treated as a parameter in MC19, which is crucial during the SN Ia explosion for the mixing between outward-expanding ash and the downward-falling cold current. This instability then creates the curvy structure along the deflagration front. These features could be frozen in the ejecta upon the expansion. However, the role of KH is unclear during the PBH ignition phase due to the very different timescale and involved process. It is possible that such instability could efficiently remove the heat from the PBH by rapidly bringing in cold fresh C+O matter. Therefore, we also present the characteristic model 06R to contrast the effect of KH instability to the progenitor and explosion dynamics.

\begin{figure}
    \centering
    \includegraphics[width=0.49\textwidth]{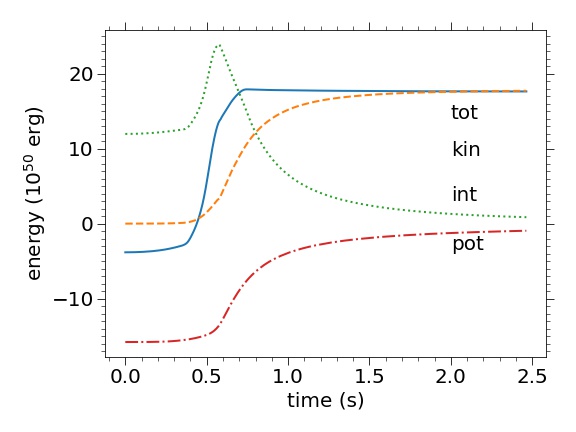}
    \includegraphics[width=0.49\textwidth]{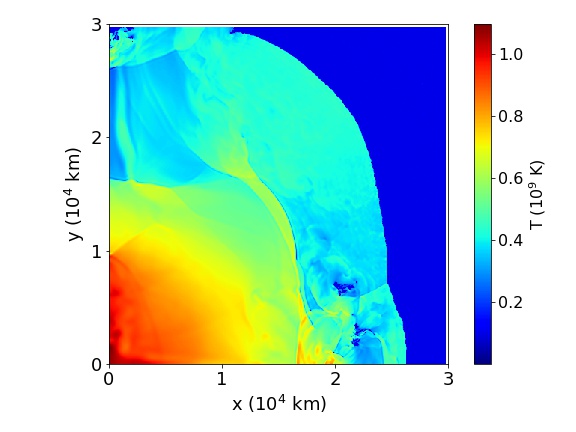}
    \caption{(top panel) The energy evolution of Model 06R for the total energy (blue solid line), kinetic energy (yellow dashed line), internal energy (green dotted line) and potential energy (red dot-dash line). 
    (bottom panel) The temperature colour-plot of the Model 06R at $\sim 1.15 s$ after the onset of explosion.}
    \label{fig:model02KH_energy}
\end{figure}

In Figure \ref{fig:model02KH_energy} we plot the energy evolution of the model. Compared to 06R-noKH, the transition to detonation occurs slightly later by $\sim 0.2$s, but the energy production is much stronger, reaching a final energy of $1.6 \times 10^{51}$ erg. Most internal energy is produced during detonation, and is  converted to kinetic energy within 2 seconds after explosion. The close-to-zero potential energy reflects that the entire star is disrupted. Comparing with 06R-noKH, the SN Ia explosion is more energetic. 

In the bottom panel we present a snapshot of the temperature colour plot of the same model at 1.15 s. The higher mass offsets the higher explosion energy and results in lower ejecta velocity, indicated by a smaller radius $\sim 3 \times 10^4$ km. The explosion structure is simpler and roughly spherical. Some less obvious shock collision trace can still be seen by the temperature discontinuity.

\begin{figure}
    \centering
    \includegraphics[width=0.49\textwidth]{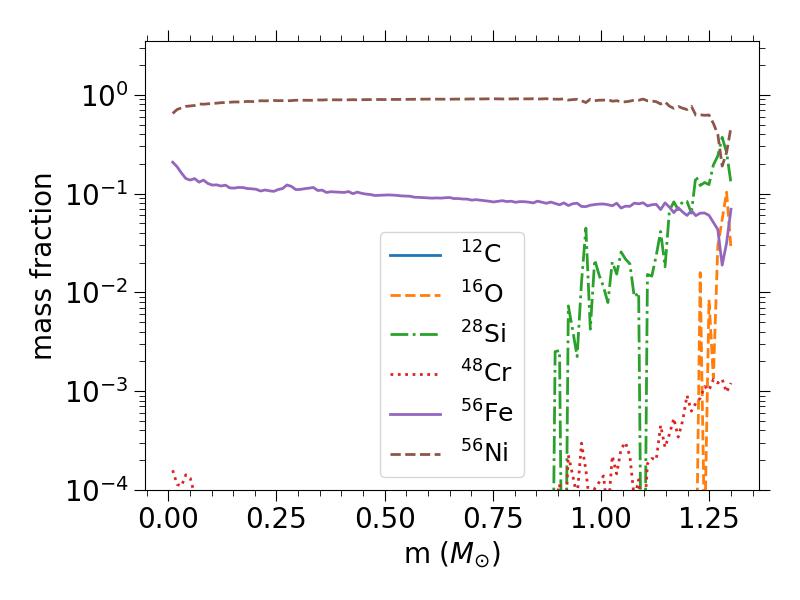}
    \includegraphics[width=0.49\textwidth]{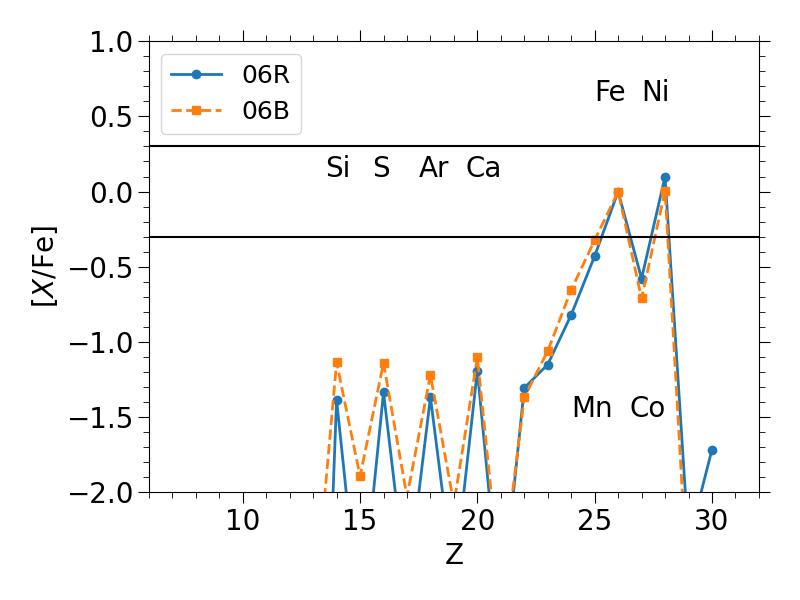}
    \caption{(top panel) The mass distribution of the ejecta for representative isotopes including $^{12}$C (blue solid line), $^{16}$O (orange dashed line), $^{28}$Si (green dot-dash line), $^{48}$Cr (red dotted line), $^{56}$Fe (purple solid line) and $^{56}$Ni (brown dashed line) in 06R.
    (bottom panel) Same as bottom panel of Figure \ref{fig:model02wKH_abund} but for 06R and 06B.}
    \label{fig:model06KH_abund}
\end{figure}

In the top panel of Figure \ref{fig:model06KH_abund} we plot the post-explosion chemical abundance profile of 06R. The star has experienced mostly complete burning and matter is made of $^{56}$Ni and $^{56}$Fe up to $\sim 1.0~M_{\odot}$ of the star. The outskirt of the star contains $^{28}$Si and a a thin layer of unburnt $^{16}$O. The chemical profile contrasts with 06R-noKH where intermediate mass elements and C+O-rich material are more prominent.

In the bottom panel, we plot the abundance ratio for 06R and 06B. Both models do not have light elements, e.g.,  C and O. The lower Si-group element production in the ejecta is highly subsolar. The typical ratios of Fe-group elements, on the other hand, are very similar to the noKH counterpart. This is expected as they are created in the core by NSE at a similar environment.
 
\section{Ignition Site as a Model Parameter}
\label{sec:dependence}

In Table \ref{tab:init_model} we list the explosion energetics and the global nucleosynthetic results of our models. The models are selected to span the ignition mass and WD mass parameter space where thermonuclear runaway is possible. 
As noted in MC19, the final form of the nuclear runaway depends on the initial orbit and the dynamical friction of the PBH. In this work, we study the two limits, bubble and ring, to explore the diversity within this mechanism. 
In two-dimensional simulations of a quadrant sphere can conveniently simulate the bubble (ring) by a semicircle along the rotation (symmetry) axis. In the column ``shape'' of the same table, we denote the initial geometry of the nuclear runaway. 

All models show successful explosions with a final energy ranging from $1.0 - 1.7 \times 10^{51}$ erg and the $^{56}$Ni mass ranging from $0.3 - 1.1 M_{\odot}$. The spread of the $^{56}$Ni agrees with the observed diversity of SNe Ia \citep{Boras2022Ni56}. The final energy of most models corresponds to strong explosions in typical SNe Ia.

The model dependence of the explosion energy differs significantly in the two series. In the KH series, the explosion properties are less sensitive to the $m-{\rm ign}$  and $M$. In the noKH series, models display stronger explosion when the mass increases, while the initial ignition being closer to the center. The results are consistent with our previous works \citep{Leung2018Chand, Leung2020SubChand} where the explosion of Ch-mass (sub-Ch mass) WD models have a lower (higher) mass dependence on progenitor mass. 

Models in the noKH series consist of mostly sub-Ch mass models where matter is less degenerate. A higher mass WD leads to higher explosion energy $E_{\rm exp}$ and more produced $^{56}$Ni. [Ni/Fe] remains very close to the solar value; while Mn remains strongly dependent on $M_{\rm WD}$. One may notice the drop of [Mn/Fe] when the ignition mass begins at a more exterior location. This is because $^{56}$Ni production is more enhanced than that of stable Mn, thus suppressing the ratio in the denominator. For 10R-noKH ad 10B-noKH, Mn production becomes significant thanks to the prominent electron capture in the degenerate matter. Models in the KH series behave like Ch-mass models where there is no significant changes in $E_{\rm exp}$, and Mn is sensitive to how close $M$ is to the Chandrasekhar mass. The Ni production is super-solar and it mildly increases with $M$.

\subsection{Explosion Dynamics}

\begin{figure}
    \centering
    \includegraphics[width=8.5cm]{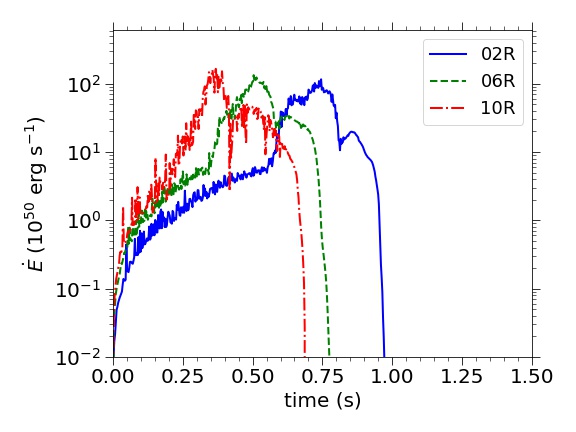}
    \includegraphics[width=8.5cm]{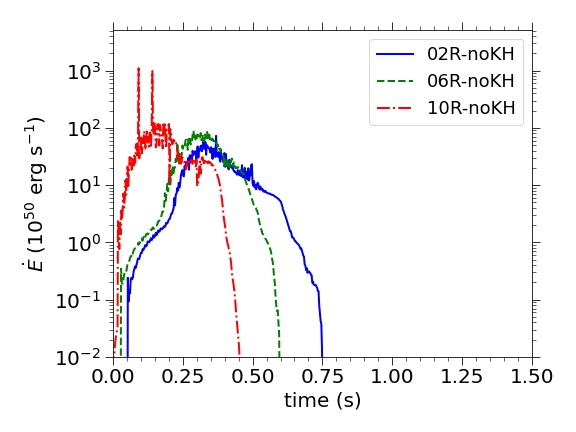}
    \caption{(top panel) The luminosity against time for the 02R (blue solid line), 06R (green dashed line) and 10R (red dotted line). (bottom panel) Same as the top panel but for the 02R-noKH (blue solid line), 06R-noKH (green dashed line), and 10R-noKH (red dotted line).}
    \label{fig:models_lumin}
\end{figure}

To further distinguish the two series, we study the explosion process and the associated nucleosynthesis. We plot in Figure \ref{fig:models_lumin} the luminosity against time for Models 02R, 06R, and 10R (upper panel) and Models 02R-noKH, 06R-noKH and 10R-noKH (lower panel). Models in the KH series share a similar peak luminosity. But models with a more interior ignition spot has a slower growth in the energy production rate. It agrees with the SN Ia models that a lower density leads to a slower deflagration and a later transition into detonation. We remind that in general the DDT transition takes place when the flame front reaches $\sim 10^7$ g cm$^{-3}$. The exact value depends on the actual turbulent strength of the deflagration flame front and the $^{12}$C burning rate. After DDT has started, it takes about 0.3 -- 0.5 seconds for detonation in all models to completely incinerate all C+O-rich matter. Models in the noKH series has an early transition and thus a shorter energy production history than the KH series. The DDT occurs very soon because the ambient density around the ignition is lower. Thus carbon detonation is dominated by the energy production. The carbon deflagration in models with a more interior ignition (e.g., 02R-noKH) lasts longer before the detonation overwhelms the energy production. We remind that in canonical SNe Ia in the binary channel, the initial runaway mostly occur near center ($< 150$ km) or near the surface by the He-C interface. 

\begin{figure}
    \centering
    \includegraphics[width=8.5cm]{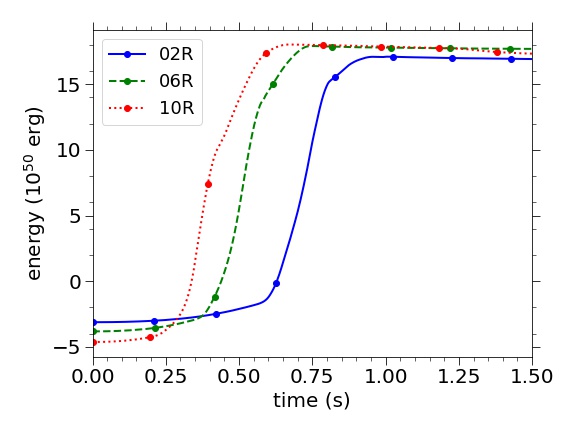}
    \includegraphics[width=8.5cm]{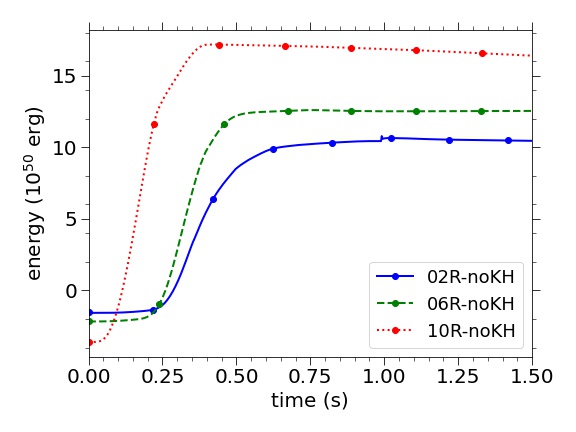}
    \caption{(top panel) The total energy against time for 02R (blue solid line), 06R (green dashed line) and 10R (red dotted line). (bottom panel) Same as the top panel but for 02R-noKH (blue solid line), 06R-noKH (green dashed line), and 10R-noKH (red dotted line).}
    \label{fig:models_energy}
\end{figure}

In Figure \ref{fig:models_energy} we plot the energy for the 02R, 06R and 10R in the upper panel and 02R-noKH, 06R-noKH, and 10R-noKH in the lower panel similar to Figure \ref{fig:models_lumin}. For the KH series, the final energy does not vary much amount models. All models have a final energy $\sim1.6 \times 10^{51}$ erg, which indicates complete burning of the WD. All exothermic reactions cease within 1 second after the initial runaway. Meanwhile, lower mass models, i.e., models with a more exterior ignition, end later. The sharp rise of energy corresponds to the detonation transition in both series, but the noKH models observe a stronger effect where the final energy reduces from $\sim 1.6 \times 10^{51}$ erg to $\sim 1.0 \times 10^{51}$ erg when the ignition mass coordinate shifts from 1.0 $M_{\odot}$ to $0.2~M_{\odot}$.

\subsection{Nucleosynthesis}

\begin{figure}
    \centering
    \includegraphics[width=8.5cm]{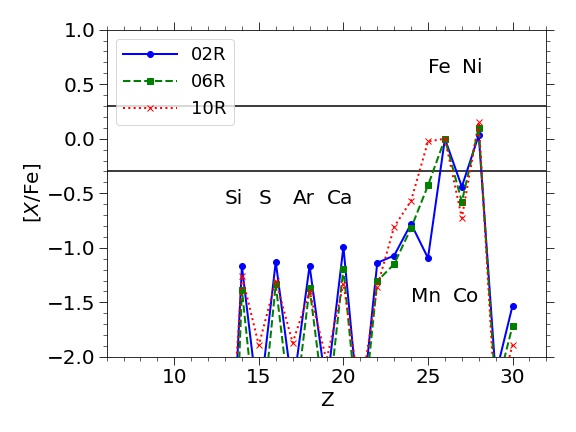}
    \includegraphics[width=8.5cm]{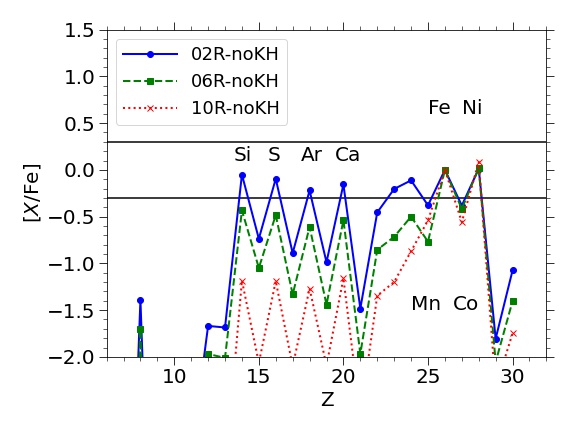}
    \caption{(top panel) The abundance ratio [Z/Fe] for 02R (blue solid line), 06R (green dashed line) and 10R (red dotted line). The two horizontal lines correspond to two times (top line) and half (bottom line) of the solar values. (bottom panel) Same as the top panel, but for 02R-noKH (blue solid line), 06R-noKH (green dashed line), and 10R-noKH (red dotted line).}
    \label{fig:models_xele}
\end{figure}

\begin{figure}
    \centering
    \includegraphics[width=8.5cm]{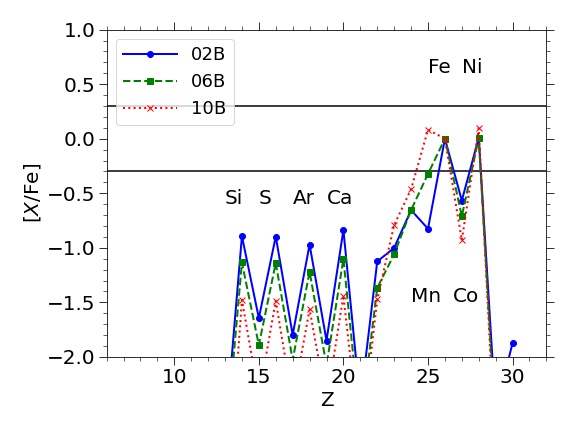}
    \includegraphics[width=8.5cm]{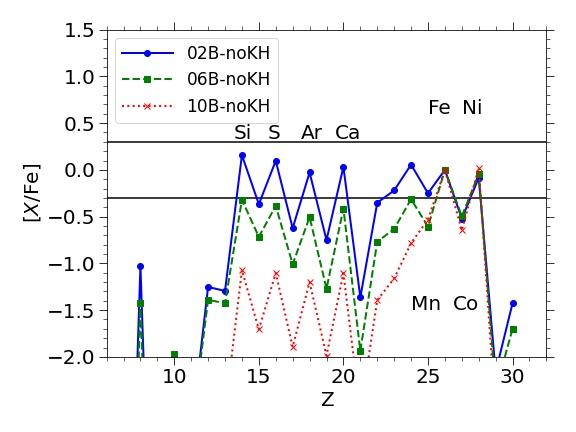}
    \caption{(top panel) Same as Figure \ref{fig:models_xele} but for 02B (blue solid line), 06B (green dashed line) and 10B (red dotted line)  (bottom panel) Same as the top panel, but for 02B-noKH (blue solid line), 06B-noKH (green dashed line), and 10B-noKH (red dotted line).}
    \label{fig:models_xele2}
\end{figure}

In Figure \ref{fig:models_xele} we show the nucleosynthetic yield of the model sequences. We plot [X/Fe] of the ejecta. Given that SNe Ia and core-collapse SNe rates are comparable in observation, in order to be compatible with the solar composition, the yield cannot be significantly higher ($\sim 2$ times) than solar values for Si- and Fe-group elements \citep{Nomoto2018SNIa}.

In all models, the full explosion disrupts the entire star. In the KH series, the models show a very similar yield that there are underproduced Si-group elements and compatible Fe-group elements with the solar abundance. The odd-even element parity is clearly seen. Notable differences appear in Cr, Mn and Co among 02R, 06R and 10R. The trends of these elements are consistent with our previous works \citep{Leung2018Chand, Leung2020SubChand}. A higher mass model substantially facilitates the production of Mn. This is because Mn can be produced mostly at low-$Y_{\rm e}$ zones through NSE ($Y_e \sim 0.46$) or through $\beta+$ decay of $^{55}$Co. This environment is typically found in high density zones due to a high Fermi energy. In contrast, a high mass model creates a smaller amount of Co. Given SNe Ia contribute to a 10 -- 50\% of the supernova rate, the underproduction of most elements suggest that this channel can only supplement the canonical (binary) explosion channel. 

In the bottom panel, we show the nucleosynthetic yield for models assuming KH instability in the progenitor. The models appears much closer to typical SN Ia models. For Model 02-woKH, most $\alpha$-chain elements yields from Si to Ni are reasonably close to the solar composition. Odd-number elements which are typically products of the electron capture, such as V, Mn and Co are also comparable with the solar abundance pattern. Similar to the top panel, a higher mass model results in a lower yield for most elements due to high Fe yield. 

In Figure \ref{fig:models_xele2} we plot the elemental ratios for the KH series (02B, 06B, and 10B) and the noKH series (02B-noKH, 06B-noKH, 10B-noKH) series. All models start with a `bubble' deflagration front. The almost identical nucleosynthetic yield suggest that the KH series cannot account for the SN Ia diversity. In the noKH series, where the explosion takes place is more important. Among all models, the 02B-noKH has the composition the closest to the solar composition, despite its sub-luminous nature. 

\subsection{Radiative Transfer}

\begin{figure}
    \centering
    \includegraphics[width=8.5cm]{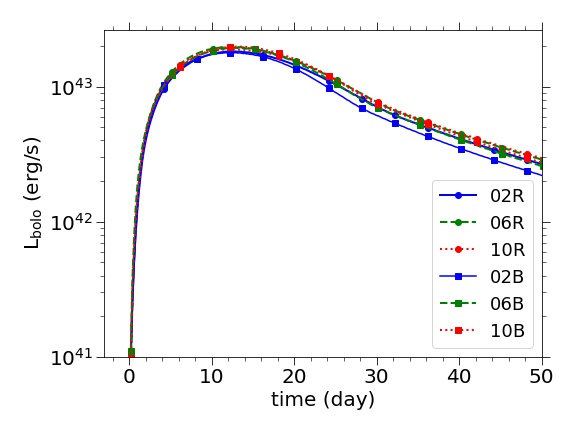}
    \includegraphics[width=8.5cm]{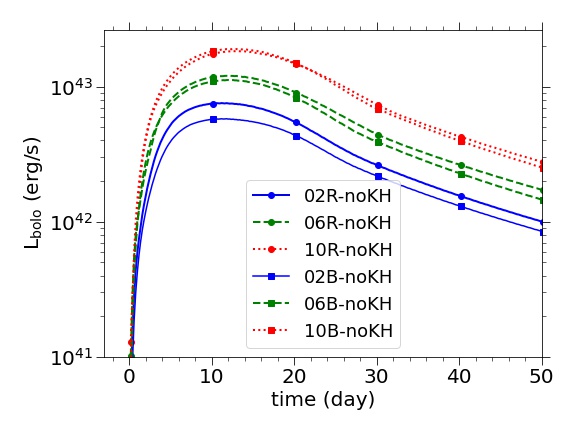}
    \caption{(top panel) The bolometric light curves of the SN Ia models in this work assuming KHI, including 02R (blue solid circles), 06R (green dashed circles), 10R (red dotted circles), 02B (blue solid squares), 06B (green dashed squares), 10B (red dotted squares). (bottom panel) Same as the top panel but for the models assuming no KH instability, including 02R-noKH (blue solid circles), 06R-noKH (green dashed circles), 10R-noKH (red dotted circles), 02B-noKH (blue solid squares), 06B-noKH (green dashed squares), 10B-noKH (red dotted squares).}
    \label{fig:models_LC}
\end{figure}

In Figure \ref{fig:models_LC} we plot in the top panel the bolometric light curves for 02R/B, 06R/B and 10R/B. The similarity in the Ch-mass like explosion leads to almost identical light curves for all three models. All models reach a peak luminosity $\sim 2\times 10^{43}$ erg s$^{-1}$ at $\sim10$ days after explosion. After that the $^{56}$Co-decay dominates the decay timescale. A mild kink is observed in all light curves, marking the transition of the ejecta turning from opaque to transparent. 

\begin{figure}
    \centering
    \includegraphics[width=8.5cm]{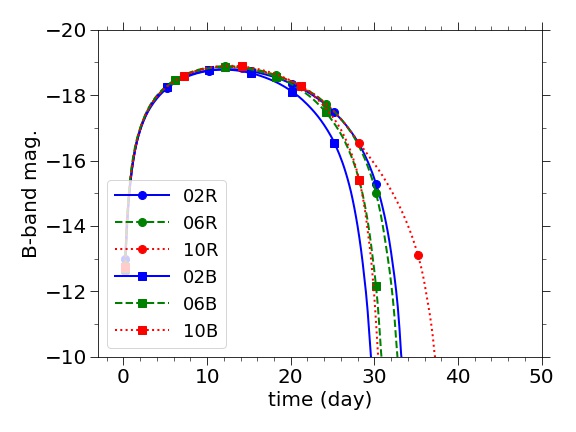}
    \caption{The B-band maximum magnitude against $\Delta M_{15}(B)$ for 02R (blue solid circles), 06R (green dashed circles) and 10R (red dotted circles), 02B (blue solid squares), 06B (green dashed squares), and 10B (red dotted squares).}
    \label{fig:models_Bband}
\end{figure}

In the bottom panel, we plot the light curves for 02-noKH, 06-noKH, and 10-noKH. The larger contrast in the progenitor mass, explosion energy and $^{56}$Ni production leads to strong distinctions in the light curves. A lower mass model results in a weaker explosion, lower peak luminosity, and a faster drop in the luminosity; meanwhile a higher mass model leads to a stronger explosion, higher peak luminosity and a slower drop in the luminosity. Such features agree with the typical SNe Ia described by the Phillip's relation that there is a self-similarity in the SN Ia light curves. 

\begin{figure}
    \centering
    \includegraphics[width=8.5cm]{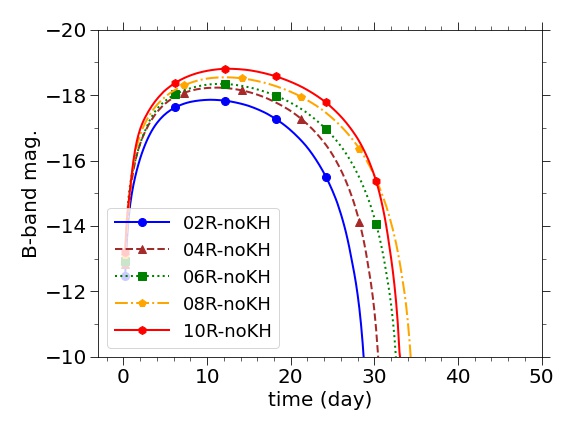}
    \includegraphics[width=8.5cm]{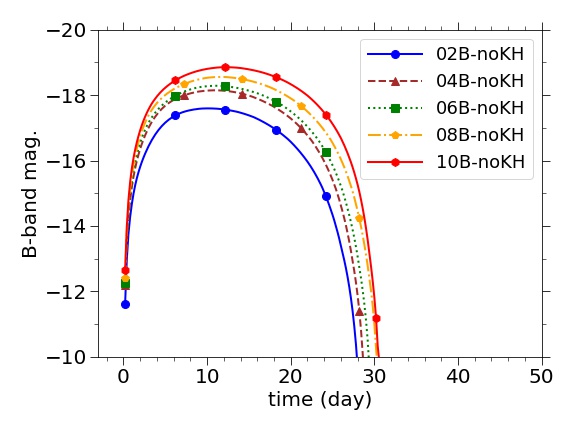}
    \caption{(top panel) Same as Figure \ref{fig:models_Bband} but for R-series assuming no KHI for 02R-noKH (blue solid circles), 04R-noKH (purple dashed triangles), 06R-noKH (green dotted squares), 08R-noKH (orange dot-dashed pentagons), 10R-noKH (red solid hexagons).
    (bottom panel) Same as the top panel but for B-series assuming no KHI, including 02B-noKH (blue solid circles), 04B-noKH (purple dashed triangles), 06B-noKH (green dotted squares), 08B-noKH (orange dot-dashed pentagons), 10B-noKH (red solid hexagons).}
    \label{fig:models_Bband2}
\end{figure}

In \texttt{SNEC}, the bolometric light curve with blackbody spectral approximation is assumed. Thus, when the luminosity profile is computed, the photosphere is located at the Lagrangian mass shell where the optical depth is 2/3. The local temperature of that mass shell is used as the effective temperature of the photosphere. Since the blackbody radiation is a sole function of temperature only, the B-band luminosity can be computed by integrating the relevant bandwidths. 

In Figure \ref{fig:models_Bband} we plot the B-band light curves for the KH series for both ignition geometry (``B'' and ``R''). The B-band is more sensitive to the ejecta as it depends on local thermodynamics quantities instead of global properties. For example, despite the bolometric light curves for 02R, 06R, 10R are very similar, differences emerge in Model 10R from the other models, showing a slower drop in the magnitude. In all models, the magnitude rapidly drops when the photosphere recedes to the center at $\sim30$ days after explosion.

In Figure \ref{fig:models_Bband2} we show the B-band for the no-KH series models. The sequential change of the light curves in the noKH series also appears in the B-band in two ways. (1) 02R-noKH, 04R-noKH, 06R-noKH and 08R-noKH show that, when mass increases, the peak magnitude increases and the magnitude decline rate decreases; (2) for Model 10R-noKH, the peak luminosity is higher but magnitude drop is faster. This is because when the model approaches the Ch-mass, the mass difference cannot offset the change in the initial ignition location. 

In the bottom panel, we plot the B-band magnitude for the ``bubble'' series. The time evolution of these models are similar to the ``R'' series. This means that the B-band is less sensitive to how the explosion is triggered, as long as the initial runaway is triggered. 

\begin{figure}
    \centering
    \includegraphics[width=8.5cm]{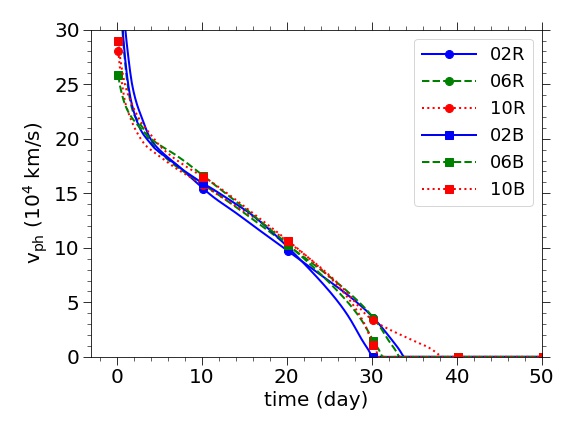}
    \includegraphics[width=8.5cm]{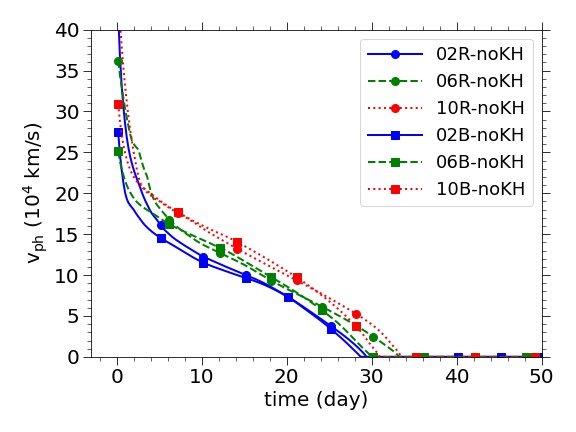}
    \caption{(top panel) The photosphere velocity of the models in the KH series, including 02R (blue solid circles), 06R (green dashed circles), 10R (red dotted circles), 02B (blue solid squares), 06B (green dashed squares), and 10B (red dotted squares). (bottom panel) Same as the top panel but for the models in the noKH series, including 02R-noKH (blue solid circles), 06R-noKH (green dashed circles), 10R-noKH (red dotted circles), 02B-noKH (blue solid squares), 06B-noKH (green dashed squares), and 10B-noKH (red dotted squares)}
    \label{fig:models_vph}
\end{figure}

In Figure \ref{fig:models_vph} we compare the photosphere velocity of selected models. The photosphere velocity can be associated with the Si- or Fe-line velocity extracted from the SN Ia spectra, which provide an addition constraint to break the degeneracy in models with similar light curves shown in \cite{Leung2021SN2019gep}. 

In the top panel, the photosphere velocity of models in the KH series are presented. In contrast to the massive star models, the three SN Ia models show an almost identical photosphere velocity. The velocity always drops without any plateau. The regression of the photosphere is shown by the steady decrease of the photosphere velocity until Day $\sim$30, where the ejecta becomes optically transparent. 

In the bottom panel, models in the noKH series are shown. The models show a minor spread in the photosphere velocity. A higher mass model results in a higher photosphere velocity by 20\% measured at the same time point. All models become transparent beyond Day $\sim$30. Despite the spread, the similarity suggest that the models will be difficult to explain SNe Ia with distinctive Si- or Fe-lines features.





\section{Discussion}
\label{sec:discussion}

\subsection{Correspondence in SN Ia Population}

\begin{figure*}
    \centering
    \includegraphics[width=17cm]{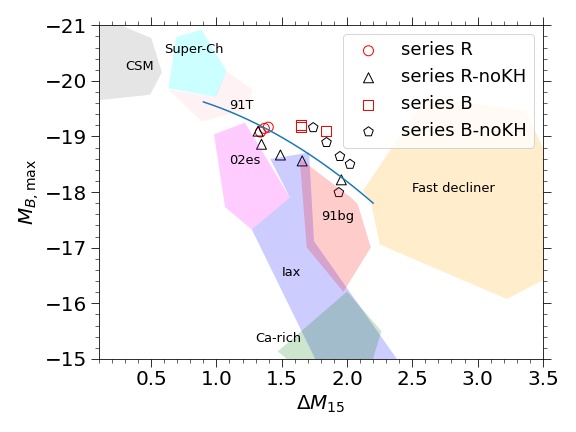}
    \caption{The $M_{B,\max}$ against $\Delta M_{15}$ for the SN Ia models assuming with (red squares) and without KH instability (black circles). The shaded areas are the parameters space outlined in \cite{Taubenberger2017SNIaDiversity} for different subtypes of SNe Ia, including SNe Ia showing CSM (CSM; grey), super-Chandrasekhar mass WD (Super-Ch; cyan), SN 1991T-like (91T; pink), SN 2002es-like (02es; magneta), SN 1991bg-like (91bg; red), Type Iax supernovae (Iax; purple), fast decliner (orange) and Calcium-rich SNe Ia (Ca-rich; green). The solid line stands for the Phillip's relation.}
    \label{fig:phillips}
\end{figure*}

SNe Ia have very diversified optical features, while simultaneously the majority following the universal Phillip's relation. Here, we compare our models with various SN subclasses to identify its possible role in the supernova family. 

In Figure \ref{fig:phillips} we plot the $M_{B,\max}$ against $\Delta M_{B,15}$ of our models. We include the classification for other subtypes of SNe Ia presented in \cite{Taubenberger2017SNIaDiversity}, qualitatively indicated by the coloured parameter spaces. They include the SNe Ia showing circumstellar medium (CSM), super-Ch-mass WD, SN 1991T-like, SN 2002es-like, SN 1991bg-like, Type Iax supernovae, fast-declining and Calcium-rich SNe Ia. The solid line represents the Phillip's relation.

Notice that while interpreting $\Delta M_{B,15}$, our models are subject to uncertainty because the magnitude is slowly changing near the peak, and rapidly falling when the photosphere recedes. The extraction of the B-band is based on the assumption of the blackbody radiation with frequency-integrated opacity. Therefore, there could be a systematic shift of our extraction of $\Delta M_{B,15}$. In plotting the figure, we use the characteristic models 06B-noKH and 06R-noKH as a reference. All models are shifted by the same amount of $\Delta M_{B,15}$ to facilitate the visualization.

The no-KH series form a sequence that agrees with the slope of the Phillip's relation closely. It overlay on the lower half where the peak magnitude between $-18 - -19$. All models can be classified as normal SNe Ia in this phase diagram. This suggests that the lateral diversity of SNe Ia around the Phillip's relation could be results of the ignition position. 

On the other hand, models from the KH series show little variations. They appear to have no difference from the brightest models without KH instability. The sub-sequence with different ignition geometry (``R'' and ``B'') behaves similarly but located at different sides of the Phillip's relation. 

\subsection{Comparison with Literature Work}

The SN Ia sequences presented in this work show a unified trend of models along the Phillips relation. This contrasts with previous works in how Ch-mass and subCh-mass models behave distinctively in the parameter space. For Ch-mass models, \cite{Ohlmann2014} showed that, by varying the central density, the model sequence changes in a direction orthogonal to the Phillips relation. For subCh-mass models, \cite{Woosley2007SubChand} showed that the simultaneous changes of $^{56}$Ni mass and ejecta mass are necessary to fit the Phillips's relation. Later work \citep[e.g., ][]{Woosley2011SubChand, Shen2021SubChand} showed that the WD progenitor mass is the primary parameter. In the violent merger scenario studied in \cite{Moll2014SubChand}, models of different masses can align with the Phillip's relation, but specific viewing angle is needed. The results are comparable with the noKH series in this work. 

A major difference of the PBH-triggered SN Ia is the reduced parameter space. For ordinary SNe Ia, the central density (progenitor mass), ignition spots and viewing angles are three independent parameters. The Phillip's relation can be realized by subsets of this multi-dimensional parameter space. In contrast, in the PBH triggered SNe Ia, the WD has a limited range of mass coordinate where the ignition can be triggered. For a given WD progenitor mass, the initial hot spot has an associated position, The rotational symmetry preserved in the bubble/ring structure further reduces the directional dependence of the viewing angle. This reduce the explosion parameter parameter to primarily the WD mass. We remind that the models still depend on other secondary parameters, e.g., the C/O ratio and the ignition details. These parameters could support the observed variety in observed SNe Ia.

While the subCh-mass models can well explain the Phillips relation, SNe Ia consisted of only subCh-mass models are insufficient owing to the nucleosynthetic perspective. Chandrasekhar mass models are the key for producing Mn in the galactic scale. The rising trend of [Mn/Fe] near [Fe/H]$\sim -1$ for stars in the Milky Way can be associated with the delay time in the Ch-mass model \citep{Kobayashi2020GCE}. Its highly degenerate matter is the natural site for electron capture and the later production of $^{55}$Co by the deflagration waves \citep{Seitenzahl2013Mn}. This environment cannot be realized in sub-Chandrasekhar mass models due to its rapid expansion after the supersonic detonation and its low matter density. In \cite{Leung2018Chand, Leung2020SubChand}, we have showed that the high density ($> 2\times 10^9 $ g cm$^{-3}$) is necessary for adequate production of Mn to match the solar composition. Some supernova remnants, e.g., 3C 397 \citep{Yamaguchi20153C397, Ohshiro2021SNR3C397} and Sagittarius A East \citep{Zhou2021SgrAE} have remarkably high ratios of [Ni/Fe] and [Mn/Fe], which leave the Ch-mass models a necessary member in the SN Ia family. 

Note that a single star can form a WD up to $\sim 1.1~M_{\odot}$. Thus, the WDs in the KH-series have progenitor mass above this threshold mass, such WDs need to experience certain period of mass accretion from its companion star, same as the binary star system in the canonical SN Ia picture. In that case, the PBH will have similar delay-time before it can be exploded, in contrast to the noKH-series.

\subsection{Future Works}

In this article, we used \texttt{SNEC} to compute the light curves with bolometric radiative transfer. While the code contains composition dependent opacity, the frequency-dependent absorption and emission is not accounted. The B-band brightness is estimated by interpolating the intensity from the perfect blackbody radiation spectra, where the local photosphere temperature is used as the reference. This method will deviate from the actual B-band brightness when the lines are densely populated in some bands. A more precise interpretation of the Phillip's relation will require multi-band radiative transfer codes, e.g., \texttt{STELLA} \citep{Blinnikov1993STELLA} and \texttt{PHOENIX} \citep{Hauschildt1999PHOENIX}. The predicted B-band brightness will account for the actual absorption and emission from individual bands, thus allowing a more robust extraction of the maximum brightness and the decline rate. 

In our simulations, we have considered the scenarios where the nuclear runaway is robustly triggered. Indeed, when the explosion occurs, the explosion timescale is much shorter than the PBH accretion timescale, where the PBH escapes from the host WD before the black hole grows and dominates the accretion process. Thus, the role of the PBH after ignition is no longer important. However, in the case where the nuclear runaway is failed, the PBH should be consistently modeled for the later evolution. This could happen in two possibilities: (1) when the PBH is accreted at a large impact parameter, thus the accreted matter has too low density to proceed efficiently carbon burning and the PBH accretes mass transfer without creating nuclear runaway; (2) when the PBH is accreted close to center, the dynamical friction could trap the PBH near the center. When the initial WD mass is close to the Ch-mass, the highly degenerate matter makes the photo-disintegration dominate energy production. As indicated in MC19, it is unclear if deflagration can be robustly occur in that scenario. The PBH could continuously consume the matter in the remnant WD and the heat from the Eddington accretion drives the WD to expand. Depending on how convection efficiency, an off-center ignition could be triggered. Such alternative possibilities are interesting extensions to be explored. 

One assumption in the formation of the first ignition is in the form of a bubble or a ring. This idealization works when the PBH triggers the explosion promptly or does not create significant internal motion. Due to the cylindrical symmetry, the orbit of the PBH defines the equatorial plane, where the initial deflagration is well approximated by a blob along the equator. However, for PBH with a high impact parameter, the PBH may require many orbits across the WD before it loses significant amount of energy by dynamical friction. The actual heated zone could create internal convection and the trajectory is spiral-like, as indicated by 3D simulations of stars engulfing brown dwarfs or planets \citep{Yarza2023PlanetEngulf, Cabezon2023DwarfPlanet}. Lastly, as noted in \cite{Niemeyer1995TurbulentFlame}, the propagation of nuclear deflagration is subject to turbulence in the form of turbulent deflagration, which are inherently three-dimensional phenomena. These observations suggest that eventually detailed three-dimensional simulations are needed to capture the physics within these objects before and during the explosion.

Finally, in the analysis of MC19, a simplified equation of state, where the specific heat does not include radiation or coulomb corrections to ions and the thermal conductivity only considers the electron-ion scattering. Due to the spherical nature, the treatment of the instabilities used empirical model to approximate the energy loss. While the additional physics will provide secondary correction to the exact parameter space of ignition mass and radius, still in order to quantitatively map out the exact relation between progenitor WD and the initial SN Ia models, 3D simulations of the ignition phase will be also necessary.

\subsection{Conclusion}

In this article we performed the first parameter study for Type Ia supernovae (SNe Ia) triggered by accretion of primordial black hole (PBH) inside C+O white dwarfs. Following \cite[MC19, ][]{Montero2019PBH}, we constructed C+O WD models in hydrostatic equilibrium for a given mass, and placed the nuclear runaway spot at selected mass coordinates where the tidal heating from the PBH could spontaneously trigger thermonuclear runaway. We studied how these WD models explode, their nucleosynthetic yields and radiative transfer. We considered the ignition scenarios where Kelvin-Helmholtz (KH) instabilities are or are not important when the first hot spot is formed. 

Models with KH instability (KH series) show clear universality among models that all models show complete explosion with most matter incinerated to form $^{56}$Ni. The explosion results in ejecta $\sim1.6 \times 10^{51}$ erg, with $\sim 1 M_{\odot}$ of $^{56}$Ni. The light curves reach $\sim2.0 \times 10^{43}$ erg s$^{-1}$ at peak. The B-band photometry shows little variations and all models appear to be super-luminous. 

Models assuming insignificant KH instability (noKH series) show a wide variation among models. The $^{56}$Ni yield ranges from $0.4-1.0 M_{\odot}$ with the final energy between $1.0 - 1.5 \times 10^{51}$ erg. The light curves have a wider spread with a peak luminosity range from $7 \times 10^{42} - 2.0 \times 10^{43}$ erg s$^{-1}$. The B-band photometry shows correlation between the peak magnitude and decline rate. 

We studied the Phillip's relation ($M_{B, \max}$ against $\Delta M_{B,15}$) of our SN Ia models. The noKH series reproduces the Phillip's relation curve approximately. The explosion resembles to ordinary SNe Ia from normal to dim ones, in the lower half of the population in the Phillip's relation. The proximity to the Phillip's relation suggests an alternative way for interpreting the universality of SNe Ia. The PBH provides a unified picture in triggering the SN Ia explosions without involving the distinctive pictures of Ch-mass and subCh-mass WDs. It also connects with the empirical Phillip's relation by a single-parameter supernova family. 

It will be interesting to study the actual capture rate of these PBH and the its contribution to the overall SN Ia statistics. This will shed light on how common this channel is compared to the canonic binary evolution scenario and clarify the origin of the empirical Phillip relation. In future, a multi-band radiative transfer will be important to uniquely pin-down the actual $M_{B, \max}$ against $\Delta M_{B,15}$ of our models. 

The trigger of PBH opens a new path for SNe Ia. The overall supernova event rates are constrained by supernova surveys such as the Sloan Digital Sky Survey \citep[SDSS, ][]{Abazajian2009SDSS7DR} and Zwicky Transient Factory \citep{Bellm2019ZTF, Dekany2020ZTF}. Further examination of the impact of PBH-triggered SNe Ia over the standard binary system channels on the overall supernova rate observed in these surveys \citep[e.g., ][]{Smith2012SDSS, Dhawan2022ZTFSNIaRate}, could shed light on the role and population of PBH as a potential candidate of DM.

\section*{Acknowledgment}

We thank Frank Timmes for the open-source subroutines of the Helmholtz equation of state and the torch nuclear reaction network. We also thank the SNEC developers for making the radiative transfer code open-source. 

This material is based upon work supported by the National Science Foundation under Grant AST-2316807.
K.N. acknowledges support by World Premier International Research Center Initiative (WPI), and JSPS KAKENHI Grant Numbers JP20K04024, JP21H04499, JP23K03452, and JP25K01046.  
S.W. thanks for the arrangement of the SUNY Polytechnic Institute Summer Undergraduate Research Program (SURP) for providing the opportunity for undergraduate research and development.   A. K. was supported by the U.S. Department of Energy (DOE) Grant No. DE-SC0009937;  by World Premier International Research Center Initiative (WPI), MEXT, Japan; and by Japan Society for the Promotion of Science (JSPS) KAKENHI Grant No. JP20H05853.

\software{  Numpy \citep{Numpy},
            Matplotlib \citep{Matplotlib},
            Pandas \citep{Pandas}
          }

\vspace{5mm}

\appendix

\section{Energy Production in WD}
\begin{figure}
    \centering
    \includegraphics[width=0.49\textwidth]{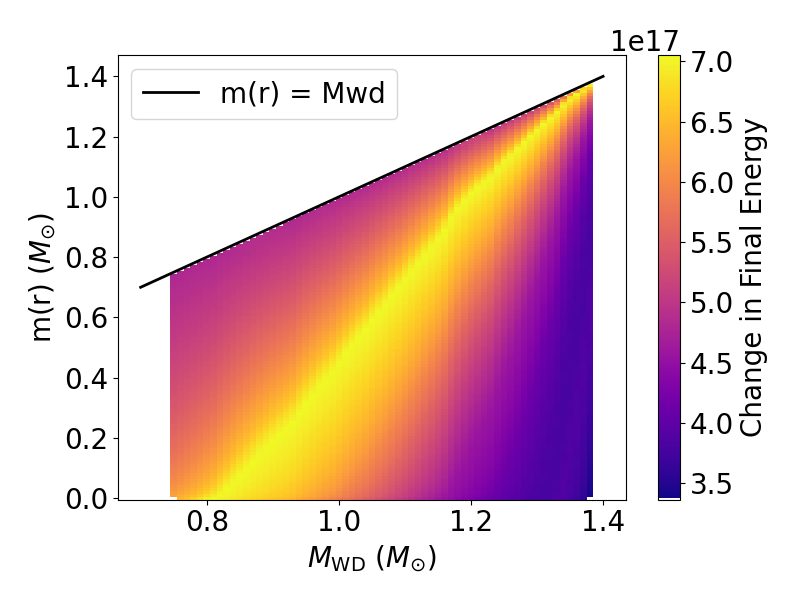}
    \includegraphics[width=0.49\textwidth]{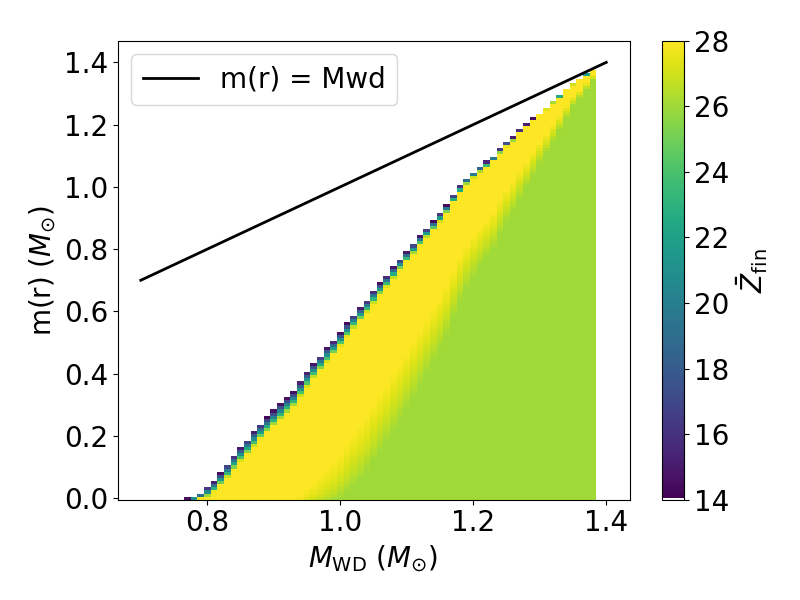}
    \caption{(left panel) The change in final energy before and after nuclear reactions at selected mass coordinates in the WD.
    (right panel) The final atomic number of the ash at given mass coordinates $m(r)$ for different progenitor WD masses $M_{\rm WD}$.}
    \label{fig:final_change_energy}
\end{figure}

In this section, we further present our estimation of the explosion based on the initial models described in Section \ref{sec:preexp}.To calculate the change in final energy at different mass coordinates $m(r)$ in the WD models, we calculate the specific internal energy before and after using the 21-isotope nuclear network ($\alpha$-chain elements with additional isotopes including $^{1}$H ,$^{3}$He, $^{4}$He, $^{14}$N, $^{54}$Fe, $^{56}$Fe, $^{56}$Cr) as outlined in \cite{Timmes1999Torch}. We use the \texttt{Helmholtz} EOS \citep{Timmes2000Helmholtz}. In the left panel of Figure \ref{fig:final_change_energy}, we show the results for the WD models with a central density from  $1 \times 10^6 - 5 \times 10^{9}$ gcm$^{-3}$. The bright yellow band corresponds to the highest energy production zone, which appears in the interior of sub-Chandrasekhar mass WD, and the outer part of high-mass WD. High mass WDs have a lower energy production because energy is consumed by the photodisintegration of $^{56}$Ni into \(\mathrm{^4He}\) nuclei. 

We also show the most abundant element by its atomic number $\bar{Z}$ in the right panel of the same figure. There is a sharp transition in three species, $^{28}$Si, $^{54}$Fe, and $^{56}$Ni. Models with $M_{\rm WD} > 1.0~M_{\odot}$ produces primarily $^{54}$Fe. The relation in the figure provides a quick estimate to the final product in PBH-triggered SNe Ia.

\section{Effects of the Initial Flame Size}

In the main text we have examined how the PBH-triggered explosions vary by its initial ignition site, and hence mass. The initial geometry of the nuclear runaway site is known to be crucial for the later dynamics in various explosion mechanisms \citep{Seitenzahl2013DDT, Lach2022GCEFlame}. In Table \ref{tab:init_model} we show that the initial ignited mass ranged from 10$^{-5}$ to 10$^{-4}~M_{\odot}$. While we unify the recipe to ignite the star, which preserves the initial size up to our resolved dimension, given that the initial ignition site is put in artificially, it questions if our explosion results depends on the initial choice of the nuclear runaway.

We run two additional models which are based on the 06R-noKH. We repeat the simulations but with an initial flame size doubled (Model 06R-noKH-L, with an initial ignition mass $1.99 \times 10^{-3}~M_{\odot}$) and the initial flame size halved (Model 06R-noKH-S, with an initial ignition mass $3.61 \times 10^{-4}~M_{\odot}$). The two models sandwich the original 06R-noKH in terms of $m_{\rm ign}$. This allows us to compare how the initial flame, which is a model parameter, affects the explosion and final nucleosynthesis.

\begin{figure*}
    \centering
    \includegraphics[width=0.49\textwidth]{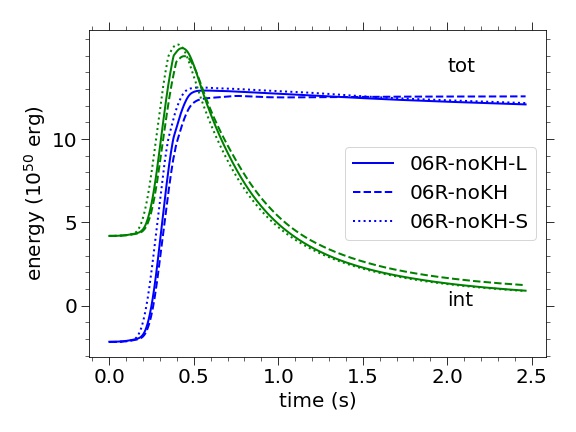}
    \includegraphics[width=0.49\textwidth]{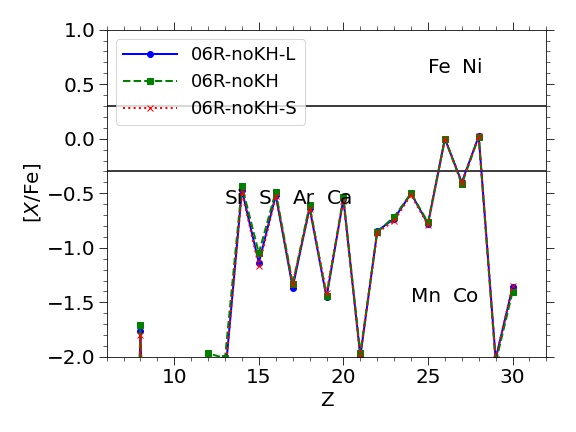}
    \caption{(left panel) The total energy (marked as tot, in blue) and the internal energy (marked as int, in green) for 06R-noKH-L (solid lines), 06R-noKH (dashed lines) and 06R-noKH-S (dotted lines) as a function of time.  
    (right panel), the final elemental abundance [X/Fe] of 06R-noKH-L (blue circles), 06R-noKH (green squares) and 06R-noKH-S (red crosses).}
    \label{fig:compare_initflame}
\end{figure*}

In the left panel of Figure \ref{fig:compare_initflame} we plot the total and internal energy. The rise of the internal energy signifies the detonation transition, where the three models agree with each other. The similarity of the shape and the magnitude of all lines shows that our explosion dynamics are insensitive to the initial configuration of the nuclear runaway. Small difference in the energy results from the escape of ejecta after the star finishes its explosion and enter homologous expansion.

In the right panel, we plot the final elemental yields of the three models after all short-lived isotopes have decayed. The three models agree very well with each other up to the size of the markers. This confirms us that the different $m_{\rm ign}$ and the minor difference in the energy production does not affect the overall nucleosynthesis, which supports our results being insensitive to the initial runaway geometry.

\bibliographystyle{aasjournal}
\pagestyle{plain}
\bibliography{biblio}

\begin{thebibliography}{}
\expandafter\ifx\csname natexlab\endcsname\relax\def\natexlab#1{#1}\fi
\providecommand{\url}[1]{\href{#1}{#1}}
\providecommand{\dodoi}[1]{doi:~\href{http://doi.org/#1}{\nolinkurl{#1}}}
\providecommand{\doeprint}[1]{\href{http://ascl.net/#1}{\nolinkurl{http://ascl.net/#1}}}
\providecommand{\doarXiv}[1]{\href{https://arxiv.org/abs/#1}{\nolinkurl{https://arxiv.org/abs/#1}}}

\bibitem[{{Abazajian} {et~al.}(2009){Abazajian}, {Adelman-McCarthy}, {Ag{\"u}eros}, {Allam}, {Allende Prieto}, {An}, {Anderson}, {Anderson}, {Annis}, {Bahcall}, {Bailer-Jones}, {Barentine}, {Bassett}, {Becker}, {Beers}, {Bell}, {Belokurov}, {Berlind}, {Berman}, {Bernardi}, {Bickerton}, {Bizyaev}, {Blakeslee}, {Blanton}, {Bochanski}, {Boroski}, {Brewington}, {Brinchmann}, {Brinkmann}, {Brunner}, {Budav{\'a}ri}, {Carey}, {Carliles}, {Carr}, {Castander}, {Cinabro}, {Connolly}, {Csabai}, {Cunha}, {Czarapata}, {Davenport}, {de Haas}, {Dilday}, {Doi}, {Eisenstein}, {Evans}, {Evans}, {Fan}, {Friedman}, {Frieman}, {Fukugita}, {G{\"a}nsicke}, {Gates}, {Gillespie}, {Gilmore}, {Gonzalez}, {Gonzalez}, {Grebel}, {Gunn}, {Gy{\"o}ry}, {Hall}, {Harding}, {Harris}, {Harvanek}, {Hawley}, {Hayes}, {Heckman}, {Hendry}, {Hennessy}, {Hindsley}, {Hoblitt}, {Hogan}, {Hogg}, {Holtzman}, {Hyde}, {Ichikawa}, {Ichikawa}, {Im}, {Ivezi{\'c}}, {Jester}, {Jiang}, {Johnson}, {Jorgensen}, {Juri{\'c}}, {Kent}, {Kessler}, {Kleinman}, {Knapp},
  {Konishi}, {Kron}, {Krzesinski}, {Kuropatkin}, {Lampeitl}, {Lebedeva}, {Lee}, {Lee}, {French Leger}, {L{\'e}pine}, {Li}, {Lima}, {Lin}, {Long}, {Loomis}, {Loveday}, {Lupton}, {Magnier}, {Malanushenko}, {Malanushenko}, {Mandelbaum}, {Margon}, {Marriner}, {Mart{\'\i}nez-Delgado}, {Matsubara}, {McGehee}, {McKay}, {Meiksin}, {Morrison}, {Mullally}, {Munn}, {Murphy}, {Nash}, {Nebot}, {Neilsen}, {Newberg}, {Newman}, {Nichol}, {Nicinski}, {Nieto-Santisteban}, {Nitta}, {Okamura}, {Oravetz}, {Ostriker}, {Owen}, {Padmanabhan}, {Pan}, {Park}, {Pauls}, {Peoples}, {Percival}, {Pier}, {Pope}, {Pourbaix}, {Price}, {Purger}, {Quinn}, {Raddick}, {Re Fiorentin}, {Richards}, {Richmond}, {Riess}, {Rix}, {Rockosi}, {Sako}, {Schlegel}, {Schneider}, {Scholz}, {Schreiber}, {Schwope}, {Seljak}, {Sesar}, {Sheldon}, {Shimasaku}, {Sibley}, {Simmons}, {Sivarani}, {Allyn Smith}, {Smith}, {Smol{\v{c}}i{\'c}}, {Snedden}, {Stebbins}, {Steinmetz}, {Stoughton}, {Strauss}, {SubbaRao}, {Suto}, {Szalay}, {Szapudi}, {Szkody}, {Tanaka},
  {Tegmark}, {Teodoro}, {Thakar}, {Tremonti}, {Tucker}, {Uomoto}, {Vanden Berk}, {Vandenberg}, {Vidrih}, {Vogeley}, {Voges}, {Vogt}, {Wadadekar}, {Watters}, {Weinberg}, {West}, {White}, {Wilhite}, {Wonders}, {Yanny}, {Yocum}, {York}, {Zehavi}, {Zibetti}, \& {Zucker}}]{Abazajian2009SDSS7DR}
{Abazajian}, K.~N., {Adelman-McCarthy}, J.~K., {Ag{\"u}eros}, M.~A., {et~al.} 2009, \apjs, 182, 543, \dodoi{10.1088/0067-0049/182/2/543}

\bibitem[{{Alcock} {et~al.}(2001){Alcock}, {Allsman}, {Alves}, {Axelrod}, {Becker}, {Bennett}, {Cook}, {Dalal}, {Drake}, {Freeman}, {Geha}, {Griest}, {Lehner}, {Marshall}, {Minniti}, {Nelson}, {Peterson}, {Popowski}, {Pratt}, {Quinn}, {Stubbs}, {Sutherland}, {Tomaney}, {Vandehei}, \& {Welch}}]{Alcock2001MACHO}
{Alcock}, C., {Allsman}, R.~A., {Alves}, D.~R., {et~al.} 2001, \apjl, 550, L169, \dodoi{10.1086/319636}

\bibitem[{{Barth} \& {Deconinck}(1999)}]{Barth1999}
{Barth}, T.~J., \& {Deconinck}, H. 1999, High-order methods for computational physics (Springer)

\bibitem[{{Bellm} {et~al.}(2019){Bellm}, {Kulkarni}, {Graham}, {Dekany}, {Smith}, {Riddle}, {Masci}, {Helou}, {Prince}, {Adams}, {Barbarino}, {Barlow}, {Bauer}, {Beck}, {Belicki}, {Biswas}, {Blagorodnova}, {Bodewits}, {Bolin}, {Brinnel}, {Brooke}, {Bue}, {Bulla}, {Burruss}, {Cenko}, {Chang}, {Connolly}, {Coughlin}, {Cromer}, {Cunningham}, {De}, {Delacroix}, {Desai}, {Duev}, {Eadie}, {Farnham}, {Feeney}, {Feindt}, {Flynn}, {Franckowiak}, {Frederick}, {Fremling}, {Gal-Yam}, {Gezari}, {Giomi}, {Goldstein}, {Golkhou}, {Goobar}, {Groom}, {Hacopians}, {Hale}, {Henning}, {Ho}, {Hover}, {Howell}, {Hung}, {Huppenkothen}, {Imel}, {Ip}, {Ivezi{\'c}}, {Jackson}, {Jones}, {Juric}, {Kasliwal}, {Kaspi}, {Kaye}, {Kelley}, {Kowalski}, {Kramer}, {Kupfer}, {Landry}, {Laher}, {Lee}, {Lin}, {Lin}, {Lunnan}, {Giomi}, {Mahabal}, {Mao}, {Miller}, {Monkewitz}, {Murphy}, {Ngeow}, {Nordin}, {Nugent}, {Ofek}, {Patterson}, {Penprase}, {Porter}, {Rauch}, {Rebbapragada}, {Reiley}, {Rigault}, {Rodriguez}, {van Roestel}, {Rusholme}, {van
  Santen}, {Schulze}, {Shupe}, {Singer}, {Soumagnac}, {Stein}, {Surace}, {Sollerman}, {Szkody}, {Taddia}, {Terek}, {Van Sistine}, {van Velzen}, {Vestrand}, {Walters}, {Ward}, {Ye}, {Yu}, {Yan}, \& {Zolkower}}]{Bellm2019ZTF}
{Bellm}, E.~C., {Kulkarni}, S.~R., {Graham}, M.~J., {et~al.} 2019, \pasp, 131, 018002, \dodoi{10.1088/1538-3873/aaecbe}

\bibitem[{{Bersten} {et~al.}(2011){Bersten}, {Benvenuto}, \& {Hamuy}}]{Bersten2011SNEC}
{Bersten}, M.~C., {Benvenuto}, O., \& {Hamuy}, M. 2011, \apj, 729, 61, \dodoi{10.1088/0004-637X/729/1/61}

\bibitem[{{Bersten} {et~al.}(2013){Bersten}, {Tanaka}, {Tominaga}, {Benvenuto}, \& {Nomoto}}]{Bersten2013SNEC}
{Bersten}, M.~C., {Tanaka}, M., {Tominaga}, N., {Benvenuto}, O.~G., \& {Nomoto}, K. 2013, \apj, 767, 143, \dodoi{10.1088/0004-637X/767/2/143}

\bibitem[{{Blinnikov} \& {Bartunov}(1993)}]{Blinnikov1993STELLA}
{Blinnikov}, S.~I., \& {Bartunov}, O.~S. 1993, \aap, 273, 106

\bibitem[{{Bora} {et~al.}(2022){Bora}, {Vink{\'o}}, \& {K{\"o}nyves-T{\'o}th}}]{Boras2022Ni56}
{Bora}, Z., {Vink{\'o}}, J., \& {K{\"o}nyves-T{\'o}th}, R. 2022, \pasp, 134, 054201, \dodoi{10.1088/1538-3873/ac63e7}

\bibitem[{{Bravo} \& {Garc{\'\i}a-Senz}(2009)}]{Bravo2009PulsatingI}
{Bravo}, E., \& {Garc{\'\i}a-Senz}, D. 2009, \apj, 695, 1244, \dodoi{10.1088/0004-637X/695/2/1244}

\bibitem[{{Bravo} {et~al.}(2009){Bravo}, {Garc{\'\i}a-Senz}, {Cabez{\'o}n}, \& {Dom{\'\i}nguez}}]{Bravo2009PulsatingII}
{Bravo}, E., {Garc{\'\i}a-Senz}, D., {Cabez{\'o}n}, R.~M., \& {Dom{\'\i}nguez}, I. 2009, \apj, 695, 1257, \dodoi{10.1088/0004-637X/695/2/1257}

\bibitem[{{Brooker} {et~al.}(2021){Brooker}, {Plewa}, \& {Fenn}}]{Brooker2021DDTCrit}
{Brooker}, E., {Plewa}, T., \& {Fenn}, D. 2021, \mnras, 501, L23, \dodoi{10.1093/mnrasl/slaa141}

\bibitem[{{Cabez{\'o}n} {et~al.}(2023){Cabez{\'o}n}, {Abia}, {Dom{\'\i}nguez}, \& {Garc{\'\i}a-Senz}}]{Cabezon2023DwarfPlanet}
{Cabez{\'o}n}, R.~M., {Abia}, C., {Dom{\'\i}nguez}, I., \& {Garc{\'\i}a-Senz}, D. 2023, \aap, 670, A155, \dodoi{10.1051/0004-6361/202244848}

\bibitem[{{Calder} {et~al.}(2007){Calder}, {Townsley}, {Seitenzahl}, {Peng}, {Messer}, {Vladimirova}, {Brown}, {Truran}, \& {Lamb}}]{Calder2007}
{Calder}, A.~C., {Townsley}, D.~M., {Seitenzahl}, I.~R., {et~al.} 2007, \apj, 656, 313, \dodoi{10.1086/510709}

\bibitem[{Carr {et~al.}(2024)Carr, Clesse, Garcia-Bellido, Hawkins, \& Kuhnel}]{Carr:2023tpt}
Carr, B., Clesse, S., Garcia-Bellido, J., Hawkins, M., \& Kuhnel, F. 2024, Phys. Rept., 1054, 1, \dodoi{10.1016/j.physrep.2023.11.005}

\bibitem[{Carr {et~al.}(2021)Carr, Clesse, Garc\'\i{}a-Bellido, \& K\"uhnel}]{Carr:2019kxo}
Carr, B., Clesse, S., Garc\'\i{}a-Bellido, J., \& K\"uhnel, F. 2021, Phys. Dark Univ., 31, 100755, \dodoi{10.1016/j.dark.2020.100755}

\bibitem[{Carr \& Kuhnel(2020)}]{Carr:2020xqk}
Carr, B., \& Kuhnel, F. 2020, Ann. Rev. Nucl. Part. Sci., 70, 355, \dodoi{10.1146/annurev-nucl-050520-125911}

\bibitem[{{Carr}(1975)}]{Carr1975}
{Carr}, B.~J. 1975, \apj, 201, 1, \dodoi{10.1086/153853}

\bibitem[{Carr \& Hawking(1974)}]{Carr:1974nx}
Carr, B.~J., \& Hawking, S.~W. 1974, Mon. Not. Roy. Astron. Soc., 168, 399

\bibitem[{Coogan {et~al.}(2021)Coogan, Morrison, \& Profumo}]{Coogan:2020tuf}
Coogan, A., Morrison, L., \& Profumo, S. 2021, Phys. Rev. Lett., 126, 171101, \dodoi{10.1103/PhysRevLett.126.171101}

\bibitem[{{Dekany} {et~al.}(2020){Dekany}, {Smith}, {Riddle}, {Feeney}, {Porter}, {Hale}, {Zolkower}, {Belicki}, {Kaye}, {Henning}, {Walters}, {Cromer}, {Delacroix}, {Rodriguez}, {Reiley}, {Mao}, {Hover}, {Murphy}, {Burruss}, {Baker}, {Kowalski}, {Reif}, {Mueller}, {Bellm}, {Graham}, \& {Kulkarni}}]{Dekany2020ZTF}
{Dekany}, R., {Smith}, R.~M., {Riddle}, R., {et~al.} 2020, \pasp, 132, 038001, \dodoi{10.1088/1538-3873/ab4ca2}

\bibitem[{{Dhawan} {et~al.}(2022){Dhawan}, {Goobar}, {Smith}, {Johansson}, {Rigault}, {Nordin}, {Biswas}, {Goldstein}, {Nugent}, {Kim}, {Miller}, {Graham}, {Medford}, {Kasliwal}, {Kulkarni}, {Duev}, {Bellm}, {Rosnet}, {Riddle}, \& {Sollerman}}]{Dhawan2022ZTFSNIaRate}
{Dhawan}, S., {Goobar}, A., {Smith}, M., {et~al.} 2022, \mnras, 510, 2228, \dodoi{10.1093/mnras/stab3093}

\bibitem[{{Fenn} {et~al.}(2016){Fenn}, {Plewa}, \& {Gawryszczak}}]{Fenn2016}
{Fenn}, D., {Plewa}, T., \& {Gawryszczak}, A. 2016, \mnras, 462, 2486, \dodoi{10.1093/mnras/stw1831}

\bibitem[{{Fink} {et~al.}(2007){Fink}, {Hillebrandt}, \& {R{\"o}pke}}]{Fink2007SubChand}
{Fink}, M., {Hillebrandt}, W., \& {R{\"o}pke}, F.~K. 2007, \aap, 476, 1133, \dodoi{10.1051/0004-6361:20078438}

\bibitem[{Flores {et~al.}(2023)Flores, Kusenko, Ghez, \& Naoz}]{Flores:2023lll}
Flores, M.~M., Kusenko, A., Ghez, A.~M., \& Naoz, S. 2023, Phys. Rev. D, 108, L061301, \dodoi{10.1103/PhysRevD.108.L061301}

\bibitem[{Fuller {et~al.}(2017)Fuller, Kusenko, \& Takhistov}]{Fuller:2017uyd}
Fuller, G.~M., Kusenko, A., \& Takhistov, V. 2017, Phys. Rev. Lett., 119, 061101, \dodoi{10.1103/PhysRevLett.119.061101}

\bibitem[{{Graham} {et~al.}(2015){Graham}, {Rajendran}, \& {Varela}}]{Graham2015DMSNIa}
{Graham}, P.~W., {Rajendran}, S., \& {Varela}, J. 2015, \prd, 92, 063007, \dodoi{10.1103/PhysRevD.92.063007}

\bibitem[{Green \& Kavanagh(2021)}]{Green:2020jor}
Green, A.~M., \& Kavanagh, B.~J. 2021, J. Phys. G, 48, 043001, \dodoi{10.1088/1361-6471/abc534}

\bibitem[{{Griest} {et~al.}(2014){Griest}, {Cieplak}, \& {Lehner}}]{Griest2014Kepler}
{Griest}, K., {Cieplak}, A.~M., \& {Lehner}, M.~J. 2014, \apj, 786, 158, \dodoi{10.1088/0004-637X/786/2/158}

\bibitem[{Harris {et~al.}(2020)Harris, Millman, van~der Walt, Gommers, Virtanen, Cournapeau, Wieser, Taylor, Berg, Smith, Kern, Picus, Hoyer, van Kerkwijk, Brett, Haldane, del R{\'{i}}o, Wiebe, Peterson, G{\'{e}}rard-Marchant, Sheppard, Reddy, Weckesser, Abbasi, Gohlke, \& Oliphant}]{Numpy}
Harris, C.~R., Millman, K.~J., van~der Walt, S.~J., {et~al.} 2020, Nature, 585, 357, \dodoi{10.1038/s41586-020-2649-2}

\bibitem[{{Hauschildt} \& {Baron}(1999)}]{Hauschildt1999PHOENIX}
{Hauschildt}, P.~H., \& {Baron}, E. 1999, Journal of Computational and Applied Mathematics, 109, 41, \dodoi{10.48550/arXiv.astro-ph/9808182}

\bibitem[{Hawking(1971)}]{Hawking:1971ei}
Hawking, S. 1971, Mon. Not. Roy. Astron. Soc., 152, 75

\bibitem[{{Hawking}(1971)}]{Hawking1971PBH}
{Hawking}, S. 1971, \mnras, 152, 75, \dodoi{10.1093/mnras/152.1.75}

\bibitem[{{Hicks}(2015)}]{Hicks2015RTFlame}
{Hicks}, E.~P. 2015, \apj, 803, 72, \dodoi{10.1088/0004-637X/803/2/72}

\bibitem[{{Hillebrandt} \& {Niemeyer}(2000)}]{Hillebrandt2000SNIaReview}
{Hillebrandt}, W., \& {Niemeyer}, J.~C. 2000, \araa, 38, 191, \dodoi{10.1146/annurev.astro.38.1.191}

\bibitem[{Hunter(2007)}]{Matplotlib}
Hunter, J.~D. 2007, Computing in Science \& Engineering, 9, 90, \dodoi{10.1109/MCSE.2007.55}

\bibitem[{{Ishigaki} {et~al.}(2021){Ishigaki}, {Hartwig}, {Tarumi}, {Leung}, {Tominaga}, {Kobayashi}, {Magg}, {Simionescu}, \& {Nomoto}}]{Ishigaki2021MPS}
{Ishigaki}, M.~N., {Hartwig}, T., {Tarumi}, Y., {et~al.} 2021, \mnras, 506, 5410, \dodoi{10.1093/mnras/stab1982}

\bibitem[{{Khokhlov}(1991)}]{Khokhlov1991DDT}
{Khokhlov}, A.~M. 1991, \aap, 245, 114

\bibitem[{{Kobayashi} {et~al.}(2020){Kobayashi}, {Leung}, \& {Nomoto}}]{Kobayashi2020GCE}
{Kobayashi}, C., {Leung}, S.-C., \& {Nomoto}, K. 2020, \apj, 895, 138, \dodoi{10.3847/1538-4357/ab8e44}

\bibitem[{{Kromer} {et~al.}(2016){Kromer}, {Fremling}, {Pakmor}, {Taubenberger}, {Amanullah}, {Cenko}, {Fransson}, {Goobar}, {Leloudas}, {Taddia}, {R{\"o}pke}, {Seitenzahl}, {Sim}, \& {Sollerman}}]{Kromer2016iPTF14atg}
{Kromer}, M., {Fremling}, C., {Pakmor}, R., {et~al.} 2016, \mnras, 459, 4428, \dodoi{10.1093/mnras/stw962}

\bibitem[{{Lach} {et~al.}(2022){Lach}, {Callan}, {Sim}, \& {R{\"o}pke}}]{Lach2022GCEFlame}
{Lach}, F., {Callan}, F.~P., {Sim}, S.~A., \& {R{\"o}pke}, F.~K. 2022, \aap, 659, A27, \dodoi{10.1051/0004-6361/202142194}

\bibitem[{{Leung} {et~al.}(2015){Leung}, {Chu}, \& {Lin}}]{Leung2015WENO}
{Leung}, S.~C., {Chu}, M.~C., \& {Lin}, L.~M. 2015, \mnras, 454, 1238, \dodoi{10.1093/mnras/stv1923}

\bibitem[{{Leung} {et~al.}(2021{\natexlab{a}}){Leung}, {Diehl}, {Nomoto}, \& {Siegert}}]{Leung2021SN2014J}
{Leung}, S.-C., {Diehl}, R., {Nomoto}, K., \& {Siegert}, T. 2021{\natexlab{a}}, \apj, 909, 152, \dodoi{10.3847/1538-4357/abc9c1}

\bibitem[{{Leung} {et~al.}(2021{\natexlab{b}}){Leung}, {Fuller}, \& {Nomoto}}]{Leung2021SN2019gep}
{Leung}, S.-C., {Fuller}, J., \& {Nomoto}, K. 2021{\natexlab{b}}, \apj, 915, 80, \dodoi{10.3847/1538-4357/abfcbe}

\bibitem[{{Leung} \& {Nomoto}(2018)}]{Leung2018Chand}
{Leung}, S.-C., \& {Nomoto}, K. 2018, \apj, 861, 143, \dodoi{10.3847/1538-4357/aac2df}

\bibitem[{{Leung} \& {Nomoto}(2020{\natexlab{a}})}]{Leung2020SubChand}
---. 2020{\natexlab{a}}, \apj, 888, 80, \dodoi{10.3847/1538-4357/ab5c1f}

\bibitem[{{Leung} \& {Nomoto}(2020{\natexlab{b}})}]{Leung2020SNIax}
---. 2020{\natexlab{b}}, \apj, 900, 54, \dodoi{10.3847/1538-4357/aba1e3}

\bibitem[{{Leung} \& {Nomoto}(2023)}]{Leung2023SNIaReview}
{Leung}, S.-C., \& {Nomoto}, K. 2023, in The Sixteenth Marcel Grossmann Meeting. On Recent Developments in Theoretical and Experimental General Relativity, Astrophysics, and Relativistic Field Theories, ed. R.~{Ruffino} \& G.~{Vereshchagin}, 4427--4446, \dodoi{10.1142/9789811269776_0374}

\bibitem[{{Liu} {et~al.}(2015){Liu}, {Moriya}, \& {Stancliffe}}]{Liu2015UVFlash}
{Liu}, Z.-W., {Moriya}, T.~J., \& {Stancliffe}, R.~J. 2015, \mnras, 454, 1192, \dodoi{10.1093/mnras/stv2076}

\bibitem[{Lu {et~al.}(2023)Lu, Picker, \& Kusenko}]{Lu:2023xoi}
Lu, Y., Picker, Z. S.~C., \& Kusenko, A. 2023.
\newblock \doarXiv{2312.15062}

\bibitem[{{Moll} {et~al.}(2014){Moll}, {Raskin}, {Kasen}, \& {Woosley}}]{Moll2014SubChand}
{Moll}, R., {Raskin}, C., {Kasen}, D., \& {Woosley}, S.~E. 2014, \apj, 785, 105, \dodoi{10.1088/0004-637X/785/2/105}

\bibitem[{{Moll} \& {Woosley}(2013)}]{Moll2013SubChand}
{Moll}, R., \& {Woosley}, S.~E. 2013, \apj, 774, 137, \dodoi{10.1088/0004-637X/774/2/137}

\bibitem[{{Montero-Camacho} {et~al.}(2019){Montero-Camacho}, {Fang}, {Vasquez}, {Silva}, \& {Hirata}}]{Montero2019PBH}
{Montero-Camacho}, P., {Fang}, X., {Vasquez}, G., {Silva}, M., \& {Hirata}, C.~M. 2019, \jcap, 2019, 031, \dodoi{10.1088/1475-7516/2019/08/031}

\bibitem[{{Morozova} {et~al.}(2015){Morozova}, {Piro}, {Renzo}, {Ott}, {Clausen}, {Couch}, {Ellis}, \& {Roberts}}]{Morozova2015SNEC}
{Morozova}, V., {Piro}, A.~L., {Renzo}, M., {et~al.} 2015, \apj, 814, 63, \dodoi{10.1088/0004-637X/814/1/63}

\bibitem[{{Niemeyer} \& {Hillebrandt}(1995)}]{Niemeyer1995TurbulentFlame}
{Niemeyer}, J.~C., \& {Hillebrandt}, W. 1995, \apj, 452, 769, \dodoi{10.1086/176345}

\bibitem[{{Niikura} {et~al.}(2019){Niikura}, {Takada}, {Yokoyama}, {Sumi}, \& {Masaki}}]{Niikura2019OGLE5}
{Niikura}, H., {Takada}, M., {Yokoyama}, S., {Sumi}, T., \& {Masaki}, S. 2019, \prd, 99, 083503, \dodoi{10.1103/PhysRevD.99.083503}

\bibitem[{Niikura {et~al.}(2019)}]{Niikura:2017zjd}
Niikura, H., {et~al.} 2019, Nature Astron., 3, 524, \dodoi{10.1038/s41550-019-0723-1}

\bibitem[{{Nomoto}(1982)}]{Nomoto1982SNIaII}
{Nomoto}, K. 1982, \apj, 257, 780, \dodoi{10.1086/160031}

\bibitem[{{Nomoto} \& {Leung}(2017)}]{Nomoto2017HBSN}
{Nomoto}, K., \& {Leung}, S.-C. 2017, in Handbook of Supernovae, ed. A.~W. {Alsabti} \& P.~{Murdin}, 1275, \dodoi{10.1007/978-3-319-21846-5_62}

\bibitem[{{Nomoto} \& {Leung}(2018)}]{Nomoto2018SNIa}
---. 2018, \ssr, 214, 67, \dodoi{10.1007/s11214-018-0499-0}

\bibitem[{{Nomoto} {et~al.}(1984){Nomoto}, {Thielemann}, \& {Yokoi}}]{Nomoto1984SNIaIII}
{Nomoto}, K., {Thielemann}, F.~K., \& {Yokoi}, K. 1984, \apj, 286, 644, \dodoi{10.1086/162639}

\bibitem[{{Ohlmann} {et~al.}(2014){Ohlmann}, {Kromer}, {Fink}, {Pakmor}, {Seitenzahl}, {Sim}, \& {R{\"o}pke}}]{Ohlmann2014}
{Ohlmann}, S.~T., {Kromer}, M., {Fink}, M., {et~al.} 2014, \aap, 572, A57, \dodoi{10.1051/0004-6361/201423924}

\bibitem[{{Ohshiro} {et~al.}(2021){Ohshiro}, {Yamaguchi}, {Leung}, {Nomoto}, {Sato}, {Tanaka}, {Okon}, {Fisher}, {Petre}, \& {Williams}}]{Ohshiro2021SNR3C397}
{Ohshiro}, Y., {Yamaguchi}, H., {Leung}, S.-C., {et~al.} 2021, \apjl, 913, L34, \dodoi{10.3847/2041-8213/abff5b}

\bibitem[{{Pakmor} {et~al.}(2012){Pakmor}, {Kromer}, {Taubenberger}, {Sim}, {R{\"o}pke}, \& {Hillebrandt}}]{Pakmor2012Merger}
{Pakmor}, R., {Kromer}, M., {Taubenberger}, S., {et~al.} 2012, \apjl, 747, L10, \dodoi{10.1088/2041-8205/747/1/L10}

\bibitem[{pandas~development team(2020)}]{Pandas}
pandas~development team, T. 2020, pandas-dev/pandas: Pandas, latest,  Zenodo, \dodoi{10.5281/zenodo.3509134}

\bibitem[{{Perlmutter} {et~al.}(1999){Perlmutter}, {Aldering}, {Goldhaber}, {Knop}, {Nugent}, {Castro}, {Deustua}, {Fabbro}, {Goobar}, {Groom}, {Hook}, {Kim}, {Kim}, {Lee}, {Nunes}, {Pain}, {Pennypacker}, {Quimby}, {Lidman}, {Ellis}, {Irwin}, {McMahon}, {Ruiz-Lapuente}, {Walton}, {Schaefer}, {Boyle}, {Filippenko}, {Matheson}, {Fruchter}, {Panagia}, {Newberg}, {Couch}, \& {Project}}]{Perlmutter1999Cosmo}
{Perlmutter}, S., {Aldering}, G., {Goldhaber}, G., {et~al.} 1999, \apj, 517, 565, \dodoi{10.1086/307221}

\bibitem[{{Phillips} {et~al.}(1999){Phillips}, {Lira}, {Suntzeff}, {Schommer}, {Hamuy}, \& {Maza}}]{Phillips1999Cosmo}
{Phillips}, M.~M., {Lira}, P., {Suntzeff}, N.~B., {et~al.} 1999, \aj, 118, 1766, \dodoi{10.1086/301032}

\bibitem[{{Plewa} {et~al.}(2004){Plewa}, {Calder}, \& {Lamb}}]{Plewa2004GCD}
{Plewa}, T., {Calder}, A.~C., \& {Lamb}, D.~Q. 2004, \apjl, 612, L37, \dodoi{10.1086/424036}

\bibitem[{{Reinecke} {et~al.}(1999{\natexlab{a}}){Reinecke}, {Hillebrandt}, \& {Niemeyer}}]{Reinecke1999SNIa}
{Reinecke}, M., {Hillebrandt}, W., \& {Niemeyer}, J.~C. 1999{\natexlab{a}}, \aap, 347, 739, \dodoi{10.48550/arXiv.astro-ph/9812120}

\bibitem[{{Reinecke} {et~al.}(2002){Reinecke}, {Hillebrandt}, \& {Niemeyer}}]{Reinecke2002SNIa}
---. 2002, \aap, 386, 936, \dodoi{10.1051/0004-6361:20020323}

\bibitem[{{Reinecke} {et~al.}(1999{\natexlab{b}}){Reinecke}, {Hillebrandt}, {Niemeyer}, {Klein}, \& {Gr{\"o}bl}}]{Reinecke1999LevelSet}
{Reinecke}, M., {Hillebrandt}, W., {Niemeyer}, J.~C., {Klein}, R., \& {Gr{\"o}bl}, A. 1999{\natexlab{b}}, \aap, 347, 724, \dodoi{10.48550/arXiv.astro-ph/9812119}

\bibitem[{{R{\"o}pke} \& {Niemeyer}(2007)}]{Roepke2007DDT}
{R{\"o}pke}, F.~K., \& {Niemeyer}, J.~C. 2007, \aap, 464, 683, \dodoi{10.1051/0004-6361:20066585}

\bibitem[{{Rosswog} {et~al.}(2009){Rosswog}, {Kasen}, {Guillochon}, \& {Ramirez-Ruiz}}]{Rosswog2009WDCollision}
{Rosswog}, S., {Kasen}, D., {Guillochon}, J., \& {Ramirez-Ruiz}, E. 2009, \apjl, 705, L128, \dodoi{10.1088/0004-637X/705/2/L128}

\bibitem[{{Seitenzahl} {et~al.}(2013{\natexlab{a}}){Seitenzahl}, {Cescutti}, {R{\"o}pke}, {Ruiter}, \& {Pakmor}}]{Seitenzahl2013Mn}
{Seitenzahl}, I.~R., {Cescutti}, G., {R{\"o}pke}, F.~K., {Ruiter}, A.~J., \& {Pakmor}, R. 2013{\natexlab{a}}, \aap, 559, L5, \dodoi{10.1051/0004-6361/201322599}

\bibitem[{{Seitenzahl} {et~al.}(2010){Seitenzahl}, {R{\"o}pke}, {Fink}, \& {Pakmor}}]{Seitenzahl2010Tracer}
{Seitenzahl}, I.~R., {R{\"o}pke}, F.~K., {Fink}, M., \& {Pakmor}, R. 2010, \mnras, 407, 2297, \dodoi{10.1111/j.1365-2966.2010.17106.x}

\bibitem[{{Seitenzahl} {et~al.}(2013{\natexlab{b}}){Seitenzahl}, {Ciaraldi-Schoolmann}, {R{\"o}pke}, {Fink}, {Hillebrandt}, {Kromer}, {Pakmor}, {Ruiter}, {Sim}, \& {Taubenberger}}]{Seitenzahl2013DDT}
{Seitenzahl}, I.~R., {Ciaraldi-Schoolmann}, F., {R{\"o}pke}, F.~K., {et~al.} 2013{\natexlab{b}}, \mnras, 429, 1156, \dodoi{10.1093/mnras/sts402}

\bibitem[{{Shen} {et~al.}(2021){Shen}, {Blondin}, {Kasen}, {Dessart}, {Townsley}, {Boos}, \& {Hillier}}]{Shen2021SubChand}
{Shen}, K.~J., {Blondin}, S., {Kasen}, D., {et~al.} 2021, \apjl, 909, L18, \dodoi{10.3847/2041-8213/abe69b}

\bibitem[{{Shields} {et~al.}(2023){Shields}, {Arunachalam}, {Kerzendorf}, {Hughes}, {Biriouk}, {Monk}, \& {Buchner}}]{Shields2023NoSNIaCompanion}
{Shields}, J.~V., {Arunachalam}, P., {Kerzendorf}, W., {et~al.} 2023, \apjl, 950, L10, \dodoi{10.3847/2041-8213/acd6a0}

\bibitem[{{Sim} {et~al.}(2010){Sim}, {R{\"o}pke}, {Hillebrandt}, {Kromer}, {Pakmor}, {Fink}, {Ruiter}, \& {Seitenzahl}}]{Sim2010SubChand}
{Sim}, S.~A., {R{\"o}pke}, F.~K., {Hillebrandt}, W., {et~al.} 2010, \apjl, 714, L52, \dodoi{10.1088/2041-8205/714/1/L52}

\bibitem[{{Sim} {et~al.}(2013){Sim}, {Seitenzahl}, {Kromer}, {Ciaraldi-Schoolmann}, {R{\"o}pke}, {Fink}, {Hillebrandt}, {Pakmor}, {Ruiter}, \& {Taubenberger}}]{Sim2013Phillips}
{Sim}, S.~A., {Seitenzahl}, I.~R., {Kromer}, M., {et~al.} 2013, \mnras, 436, 333, \dodoi{10.1093/mnras/stt1574}

\bibitem[{{Smith} {et~al.}(2012){Smith}, {Nichol}, {Dilday}, {Marriner}, {Kessler}, {Bassett}, {Cinabro}, {Frieman}, {Garnavich}, {Jha}, {Lampeitl}, {Sako}, {Schneider}, \& {Sollerman}}]{Smith2012SDSS}
{Smith}, M., {Nichol}, R.~C., {Dilday}, B., {et~al.} 2012, \apj, 755, 61, \dodoi{10.1088/0004-637X/755/1/61}

\bibitem[{Sugiyama {et~al.}(2023)Sugiyama, Takada, \& Kusenko}]{Sugiyama:2021xqg}
Sugiyama, S., Takada, M., \& Kusenko, A. 2023, Phys. Lett. B, 840, 137891, \dodoi{10.1016/j.physletb.2023.137891}

\bibitem[{Takhistov {et~al.}(2021)Takhistov, Fuller, \& Kusenko}]{Takhistov:2020vxs}
Takhistov, V., Fuller, G.~M., \& Kusenko, A. 2021, Phys. Rev. Lett., 126, 071101, \dodoi{10.1103/PhysRevLett.126.071101}

\bibitem[{{Tanikawa} {et~al.}(2019){Tanikawa}, {Nomoto}, {Nakasato}, \& {Maeda}}]{Tanikawa2019Merger}
{Tanikawa}, A., {Nomoto}, K., {Nakasato}, N., \& {Maeda}, K. 2019, \apj, 885, 103, \dodoi{10.3847/1538-4357/ab46b6}

\bibitem[{{Taubenberger}(2017)}]{Taubenberger2017SNIaDiversity}
{Taubenberger}, S. 2017, in Handbook of Supernovae, ed. A.~W. {Alsabti} \& P.~{Murdin}, 317, \dodoi{10.1007/978-3-319-21846-5_37}

\bibitem[{{Timmes}(1999)}]{Timmes1999Torch}
{Timmes}, F.~X. 1999, \apjs, 124, 241, \dodoi{10.1086/313257}

\bibitem[{{Timmes} \& {Arnett}(1999)}]{Timmes1999Helmholtz}
{Timmes}, F.~X., \& {Arnett}, D. 1999, \apjs, 125, 277, \dodoi{10.1086/313271}

\bibitem[{{Timmes} {et~al.}(2000){Timmes}, {Hoffman}, \& {Woosley}}]{Timmes20007Iso}
{Timmes}, F.~X., {Hoffman}, R.~D., \& {Woosley}, S.~E. 2000, \apjs, 129, 377, \dodoi{10.1086/313407}

\bibitem[{{Timmes} \& {Swesty}(2000)}]{Timmes2000Helmholtz}
{Timmes}, F.~X., \& {Swesty}, F.~D. 2000, \apjs, 126, 501, \dodoi{10.1086/313304}

\bibitem[{{Timmes} \& {Woosley}(1992)}]{Timmes1992COFLame}
{Timmes}, F.~X., \& {Woosley}, S.~E. 1992, \apj, 396, 649, \dodoi{10.1086/171746}

\bibitem[{{Tisserand} {et~al.}(2007){Tisserand}, {Le Guillou}, {Afonso}, {Albert}, {Andersen}, {Ansari}, {Aubourg}, {Bareyre}, {Beaulieu}, {Charlot}, {Coutures}, {Ferlet}, {Fouqu{\'e}}, {Glicenstein}, {Goldman}, {Gould}, {Graff}, {Gros}, {Haissinski}, {Hamadache}, {de Kat}, {Lasserre}, {Lesquoy}, {Loup}, {Magneville}, {Marquette}, {Maurice}, {Maury}, {Milsztajn}, {Moniez}, {Palanque-Delabrouille}, {Perdereau}, {Rahal}, {Rich}, {Spiro}, {Vidal-Madjar}, {Vigroux}, {Zylberajch}, \& {EROS-2 Collaboration}}]{Tisserand2007EROS}
{Tisserand}, P., {Le Guillou}, L., {Afonso}, C., {et~al.} 2007, \aap, 469, 387, \dodoi{10.1051/0004-6361:20066017}

\bibitem[{{Tiwari} {et~al.}(2022){Tiwari}, {Graur}, {Fisher}, {Seitenzahl}, {Leung}, {Nomoto}, {Perets}, \& {Shen}}]{Tiwari2022SNIaLC}
{Tiwari}, V., {Graur}, O., {Fisher}, R., {et~al.} 2022, \mnras, 515, 3703, \dodoi{10.1093/mnras/stac1618}

\bibitem[{{Travaglio} {et~al.}(2004){Travaglio}, {Hillebrandt}, {Reinecke}, \& {Thielemann}}]{Travaglio2004Tracer}
{Travaglio}, C., {Hillebrandt}, W., {Reinecke}, M., \& {Thielemann}, F.~K. 2004, \aap, 425, 1029, \dodoi{10.1051/0004-6361:20041108}

\bibitem[{Wang \& Spiteri(2007)}]{Wang2007WENO}
Wang, R., \& Spiteri, R.~J. 2007, SIAM Journal on Numerical Analysis, 45, 1871, \dodoi{10.1137/050637868}

\bibitem[{{Weng} {et~al.}(2022){Weng}, {Zhou}, {Chen}, {Leung}, {Toonen}, {Perets}, {Nomoto}, {Zenati}, \& {Vink}}]{Weng2022SNR}
{Weng}, J., {Zhou}, P., {Chen}, Y., {et~al.} 2022, \apj, 924, 119, \dodoi{10.3847/1538-4357/ac308d}

\bibitem[{{Woosley} \& {Kasen}(2011)}]{Woosley2011SubChand}
{Woosley}, S.~E., \& {Kasen}, D. 2011, \apj, 734, 38, \dodoi{10.1088/0004-637X/734/1/38}

\bibitem[{{Woosley} {et~al.}(2007){Woosley}, {Kasen}, {Blinnikov}, \& {Sorokina}}]{Woosley2007SubChand}
{Woosley}, S.~E., {Kasen}, D., {Blinnikov}, S., \& {Sorokina}, E. 2007, \apj, 662, 487, \dodoi{10.1086/513732}

\bibitem[{{Yamaguchi} {et~al.}(2015){Yamaguchi}, {Badenes}, {Foster}, {Bravo}, {Williams}, {Maeda}, {Nobukawa}, {Eriksen}, {Brickhouse}, {Petre}, \& {Koyama}}]{Yamaguchi20153C397}
{Yamaguchi}, H., {Badenes}, C., {Foster}, A.~R., {et~al.} 2015, \apjl, 801, L31, \dodoi{10.1088/2041-8205/801/2/L31}

\bibitem[{{Yarza} {et~al.}(2023){Yarza}, {Razo-L{\'o}pez}, {Murguia-Berthier}, {Everson}, {Antoni}, {MacLeod}, {Soares-Furtado}, {Lee}, \& {Ramirez-Ruiz}}]{Yarza2023PlanetEngulf}
{Yarza}, R., {Razo-L{\'o}pez}, N.~B., {Murguia-Berthier}, A., {et~al.} 2023, \apj, 954, 176, \dodoi{10.3847/1538-4357/acbdfc}

\bibitem[{{Zel'dovich} \& {Novikov}(1967)}]{Zeldovich:1967}
{Zel'dovich}, Y.~B., \& {Novikov}, I.~D. 1967, Sov. Astron., 10, 602

\bibitem[{{Zhou} {et~al.}(2021){Zhou}, {Leung}, {Li}, {Nomoto}, {Vink}, \& {Chen}}]{Zhou2021SgrAE}
{Zhou}, P., {Leung}, S.-C., {Li}, Z., {et~al.} 2021, \apj, 908, 31, \dodoi{10.3847/1538-4357/abbd45}

\end{thebibliography}

\end{document}